\documentclass[letterpaper,showpacs,preprintnumbers,amsmath,amssymb,nofootinbib,
twocolumn,
superscriptaddress,notitlepage,prd]{revtex4-1}

    \makeatletter
\def\@fnsymbol#1{\ensuremath{\ifcase#1\or \dagger\or \ddagger\or
   \mathsection\or \mathparagraph\or \|\or **\or \dagger\dagger
   \or \ddagger\ddagger \else\@ctrerr\fi}}
    \makeatother

\usepackage{graphicx}
\usepackage[usenames,dvipsnames]{color}
\usepackage[
colorlinks=true,       
linkcolor=red,          
citecolor=blue,        
filecolor=magenta,      
urlcolor=cyan  
]{hyperref}

\hypersetup{colorlinks,citecolor= blue,linkcolor= blue, urlcolor=blue}

\usepackage{amsmath}
\usepackage{bbm}
\usepackage{amsfonts}
\usepackage{amssymb}
\usepackage{latexsym}
\usepackage{graphicx}
\usepackage[english]{babel}
\usepackage{multirow}
\usepackage{float}
\usepackage{url}
\usepackage{slashed}
\usepackage{xcolor} 
\usepackage{array}
\usepackage{cases}

\newcommand{\be}{\begin{equation}}
\newcommand{\ee}{\end{equation}}
\newcommand{\ba}{\begin{array}}
\newcommand{\ea}{\end{array}}
\newcommand{\bea}{\begin{eqnarray}}
\newcommand{\eea}{\end{eqnarray}}

\usepackage{ulem,fancyvrb}
\usepackage{xcolor}

\begin{document}

\title{Enhanced long-lived dark photon signals at lifetime frontier detectors}

\author{Mingxuan Du} 
\affiliation{Department of Physics, Nanjing University, Nanjing 210093, China}

\author{Rundong Fang} 
\affiliation{Department of Physics, Nanjing University, Nanjing 210093, China}

\author{Zuowei Liu} 
\affiliation{Department of Physics, Nanjing University, Nanjing 210093, China} 
\affiliation{CAS Center for Excellence in Particle Physics, Beijing 100049, China}

\author{Van Que Tran} \thanks{Present address: Tsung-Dao Lee Institute, Shanghai Jiao Tong University, 520 Shengrong Road, Shanghai 200240, China}
\affiliation{Department of Physics, Nanjing University, Nanjing 210093, China}
\affiliation{Faculty of Fundamental Sciences, PHENIKAA University, Yen Nghia, Ha Dong, Hanoi 12116, Vietnam}

\begin{abstract}

Long-lived particles that are present in many new physics models beyond the standard model, can be searched for in a number of newly proposed lifetime frontier experiments at the LHC. The signals of the long-lived dark photons can be significantly enhanced in a new dark photon model in which dark photons are copiously produced in the hidden radiation process. We investigate the capability of various lifetime frontier detectors in probing the parameter space of this model, including the far forward detectors FACET and FASER, the far transverse detector MATHUSLA, and the precision timing detector CMS-MTD. We find that the accessible parameter space is significantly enlarged by the hidden radiation process so that FACET, MATHUSLA, and CMS-MTD can probe a much larger parameter space than the so-called minimal model. The parameter space probed by FACET is found to be much larger than FASER, which is largely due to the fact that the former has a larger decay volume and is closer to the interaction point. There also exists some parameter space that can be probed both by the far detectors and by precision timing detectors, so that different experiments can be complementary to each other. A brief overview of the lifetime frontier detectors is also given.

\end{abstract}

\maketitle

\section{introduction}

Particles with a macroscopic decay length, 
ranging from a few {centimeters}
to several hundred 
meters and beyond, can be classified as long-lived particles (LLPs) 
at the large hadron collider 
(LHC). 
Such LLPs are endemic in new physics models 
beyond the standard model (SM); 
see e.g.\ 
\cite{Lee:2018pag, Alimena:2019zri} for recent reviews. 
A number of new detectors at the LHC 
have been recently proposed to 
search for LLPs, which can be collectively referred to 
as lifetime frontier detectors. 
These include the detectors that are placed in the forward region: 
FACET~\cite{FACET:talk1, Cerci:2021nlb}, 
FASER~\cite{Feng:2017uoz, Ariga:2018zuc, Ariga:2018uku, Ariga:2018pin, Ariga:2019ufm}, 
{FASER2} \cite{ Ariga:2019ufm, Kling:2021fwx}, 
AL3X~\cite{Gligorov:2018vkc}, 
and {MoEDAL-MAPP} \cite{Staelens:2019gzt}; 
the detectors that are placed in the central region:
MATHUSLA 
\cite{Chou:2016lxi, Curtin:2018mvb, Alpigiani:2018fgd, Lubatti:2019vkf, Alpigiani:2020tva}, 
CODEX-b~\cite{Gligorov:2017nwh, Aielli:2019ivi}, 
ANUBIS~\cite{Bauer:2019vqk}; 
and the precision timing detectors that are to be installed 
at ATLAS, CMS, and LHCb to mitigate the pileup backgrounds 
in the coming HL-LHC phase: 
CMS-MTD~\cite{CMStiming}, 
ATLAS-HGTD~\cite{Allaire:2018bof}, 
LHCb-TORCH~\cite{LHCb:2017MTD, LHCb:2008vvz}. 
A plethora of LLPs can be studied in the newly proposed 
lifetime frontier detectors 
\cite{Curtin:2017izq, 
Evans:2017lvd,
Feng:2017vli, 
Kling:2018wct, 
Helo:2018qej,
Liu:2018wte, 
Feng:2018pew,
Cerri:2018rkm, 
Curtin:2018ees,
Berlin:2018jbm, 
Dercks:2018eua, 
Dercks:2018wum,
Deppisch:2018eth,
Flowers:2019gvj, 
Kim:2019oyh,
Mason:2019okp, 
Boiarska:2019vid,
No:2019gvl, 
Krovi:2019hdl,
Jodlowski:2019ycu,
Du:2019mlc,
Deppisch:2019kvs,
Hirsch:2020klk, 
Yuan:2020eeu,
Liu:2020vur,
Dreiner:2020qbi, 
DeVries:2020jbs,
Bertuzzo:2020rzo,
Takahashi:2021tff,
Cottin:2021lzz, 
Cheung:2021utb,
Guo:2021vpb,
Bhattacherjee:2021rml,
Mitsou:2021tti,
Das:2021nqj}.

One well-motivated new physics particle is the dark photon 
(denoted by $A'_\mu$)
which can naturally arise
in kinetic mixing model 
\cite{Holdom:1985ag, 
Foot:1991kb}, 
in Stueckelberg models 
\cite{Kors:2005uz,
Feldman:2006ce, 
Feldman:2006wb, 
Feldman:2007wj, 
Feldman:2009wv} 
\cite{Du:2019mlc}.
The interaction between  
the dark photon $A'_\mu$ and the SM fermion $f$ can be parametrized 
as
\be 
e \, \epsilon \, Q_f \, A'_\mu \bar f \gamma^\mu f. 
\label{eq:epsilon}
\ee 
Long-lived dark photons (LLDPs) have a small $\epsilon$ coupling, 
which, however, leads to a suppressed collider signal.  
Recently, a new dark photon model is proposed in Ref.\ \cite{Du:2019mlc} 
where the dark photon is produced at colliders by the hidden fermion 
radiation so that the collider signal no longer suffers from the small 
$\epsilon$ parameter. For that reason, the LLDP signal at the LHC 
in this new dark photon model can be significantly enhanced.\footnote{See 
\cite{Buschmann:2015awa, Arguelles:2016ney, Kim:2019oyh, Krovi:2019hdl} 
for other dark photon models with a sizeable LLDP signal.}
Thus, we will refer to the dark photon models,    
where the dark photon interacts with the SM sector only via 
the interaction Lagrangian in Eq.\ \eqref{eq:epsilon}, 
as the ``minimal'' dark photon models, to be distinguished 
from the dark photon models proposed in 
Ref.\ \cite{Du:2019mlc}.

In this paper, we investigate the capability of various lifetime frontier 
detectors in probing the parameter space of LLDPs both 
in the minimal dark photon model
and in the newly proposed dark photon model \cite{Du:2019mlc}. 
We carry out detailed analysis for detectors: 
the far forward detector, FACET and FASER, 
the central transverse detector, MATHUSLA, 
and the precision timing detector, CMS-MTD. 
We compute   
the expected limits from these detectors. 
We find that the parameter space probed by 
FACET and MATHUSLA
are significantly enlarged
by the hidden fermion radiation in the new dark photon model, 
as compared to the minimal dark photon model. 
We also find that the LLDP signal at the newly proposed 
far detector FACET is significantly larger 
than FASER, owing to a larger decay volume 
and a shorter distance to the interaction point of the FACET detector.

The rest of the paper is organized as follows. 
We briefly review the dark photon model that has 
an enhanced LLDP signal 
in section~\ref{sec:model}.
A mini-overview on lifetime-frontier detectors is given 
in section~\ref{sec: detector-review}.
We discuss three main DP production channels 
in section~\ref{sec:DP_production}.
We analyze the signal events in different lifetime-frontier detectors 
in section~\ref{sec:simu-and-considerations}. 
Given 
in section~\ref{sec:result} 
are the {sensitivities} to the parameter space from four different detectors: 
FACET, 
FASER(2), 
MATHUSLA, 
and CMS-MTD. 
A semi-analytic comparison between far detectors is given 
in section~\ref{sec:facet-faser-comparison}.
We summarize our findings in section~\ref{sec:summary}.

\section{The model and its parameter space}
\label{sec:model}

In this analysis, we consider 
the dark photon model 
that has been proposed 
recently to enhance the (suppressed) long-lived 
dark photon signal at the LHC \cite{Du:2019mlc}.
In this model, 
the standard model is extended by a hidden sector 
that consists of two Abelian gauge groups 
$U(1)_F$ and $U(1)_W$
with corresponding gauge bosons 
$X_\mu$ and $C_\mu$ respectively, and 
one Dirac fermion $\psi$ charged under both gauge 
groups \cite{Du:2019mlc}. 
The gauge boson mass terms (due to the Stueckelberg 
mechanism 
\cite{Kors:2005uz, 
Feldman:2006ce, 
Feldman:2006wb, 
Feldman:2007wj, 
Feldman:2009wv}) 
and the gauge interaction terms in 
the hidden sector are given by
\bea
\label{eq:lagrangian}
{\cal L} = 
- \frac{1}{2} ( \partial_\mu \sigma_1 +  m_{1}\epsilon_1 B_{\mu} +  m_{1} X_{\mu} )^2 \nonumber \\
- \frac{1}{2} ( \partial_\mu \sigma_2 +  m_{2}\epsilon_2 B_{\mu} +  m_{2} C_{\mu} )^2  \nonumber \\
 + g_F \bar \psi \gamma^\mu \psi X_{\mu} 
+ g_W \bar \psi \gamma^\mu \psi C_{\mu}, 
\eea
where $B_\mu$ is the hypercharge boson in the SM, 
$\sigma_1$ and $\sigma_2$ are the axion fields in the Stueckelberg 
mechanism,
$g_F$ and $g_W$ are the gauge coupling constants, 
and $m_1$, $m_2$, $m_{1}\epsilon_1$, 
and $m_{2}\epsilon_2$ are mass terms in 
the Stueckelberg mechanism 
with $\epsilon_{1,2}$ being (small) 
dimensionless numbers.

The 2 by 2 neutral gauge boson mass matrix {in the SM} 
is
extended to a 4 by 4 mass matrix due to the fact that 
the two new gauge bosons, $X_\mu$ and $C_\mu$, 
have mixed mass terms with the SM hypercharge boson $B_\mu$; 
the new neutral gauge boson mass matrix 
in the basis of $V= ( C,X, B, A^3)$
is given by \cite{Du:2019mlc}
\be
M^2 = 
\begin{pmatrix} 
m_{2}^2  & 0 & m_{2}^2 \epsilon_2 & 0\cr
0 & m_{1}^2 & m_{1}^2 \epsilon_1 & 0 \cr
m_{2}^2 \epsilon_2 & m_{1}^2 \epsilon_1 &  
\sum\limits_{i=1}^2 m_{i}^2 \epsilon_i^2 + {g'^2 v^2 \over 4} 
  & - {g'g v^2 \over 4}  \cr
0 &  0 & - {g'g v^2 \over 4} & {g^2 v^2 \over 4}
\end{pmatrix} 
\label{eq:massmatrix}
\ee
where $A^3$ is the third component of the $SU(2)_L$ gauge 
bosons, $g$ and $g'$ are gauge couplings for 
the SM $SU(2)_L$ and $U(1)_Y$ 
gauge groups respectively,  
and $v = 246$ GeV is the vacuum expectation value of the SM Higgs boson.

Diagonalization of the mass matrix (via an orthogonal transformation
${\cal O}$) leads to the mass eigenstates 
$E= ( Z', A', Z, A)$  with $E_i={\cal O}_{ji} V_{j}$ 
where $A$ is the SM photon, 
$Z$ is the SM $Z$ boson, 
$A'$ is the dark photon, 
and $Z'$ is the new heavy boson.
The interaction Lagrangian between 
the mass eigenstates of the neutral gauge {bosons}
and the fermions is given by \cite{Du:2019mlc} 
\be
\label{eq:coupling}
\left[ \bar f \gamma_\mu (v^f_i - \gamma_5 a^f_i) f 
+ v^\psi_i \bar \psi \gamma_\mu  \psi  \right] E^\mu_i 
\ee
where $f$ is the SM fermion. 
The small coupling $v_4^\psi$ between the hidden 
fermion $\psi$ and the SM photon can be 
rewritten as  $v_4^\psi \equiv e \delta$ 
where $\delta$ is usually referred to as 
``millicharge''.

\begin{figure}[htbp]
\begin{centering}
\includegraphics[width=0.4\textwidth]{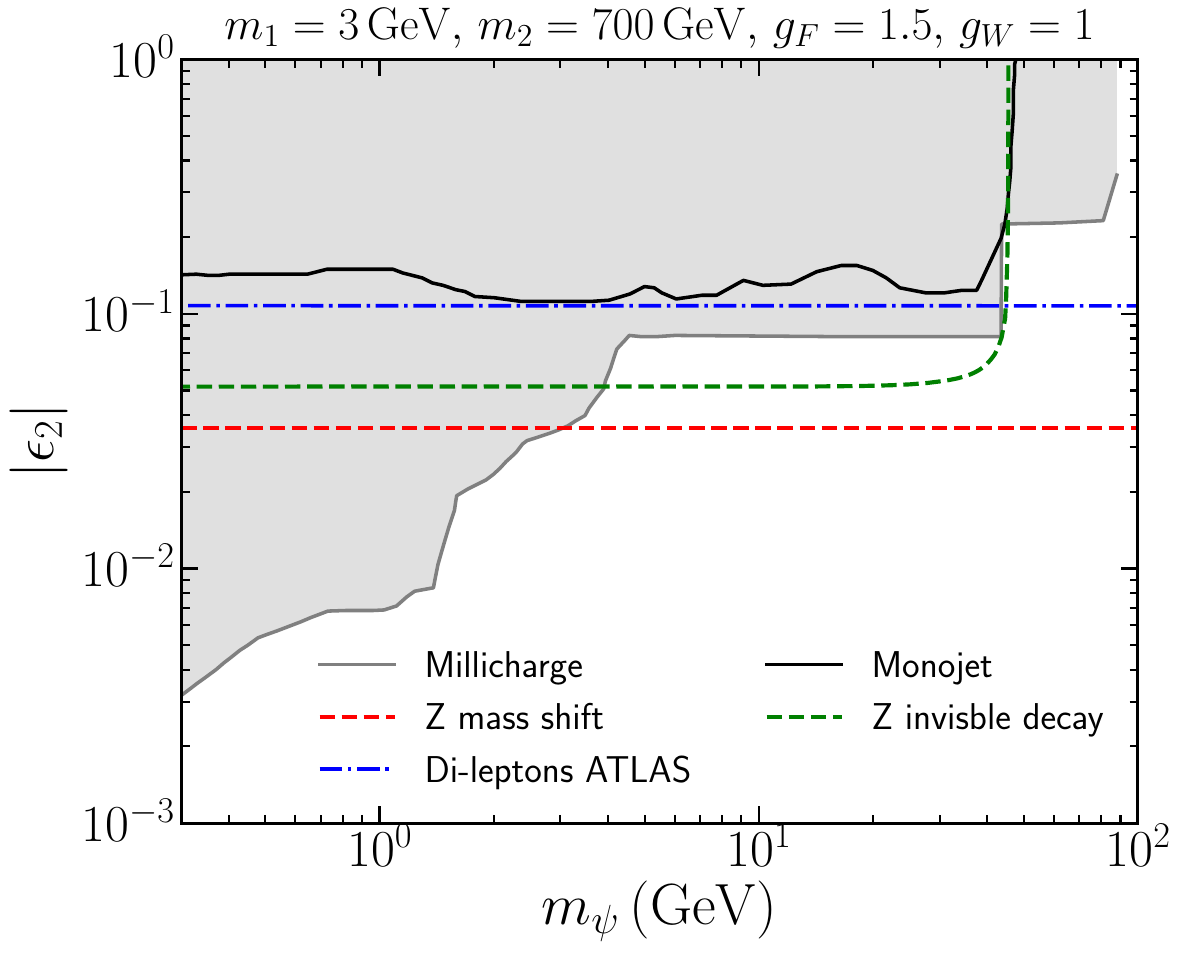}
\caption{The upper bound on $\epsilon_2$ as a function of $m_\psi$. 
The other parameters are  
$m_1 =  3$ GeV, $m_2 =  700$ GeV, 
$g_F = 1.5$, $g_W = 1.0$, 
and $\epsilon_1 \ll \epsilon_2$. 
Here $\epsilon_2 \simeq (-g'/g_W) \delta$ where $\delta$ is the millicharge 
of $\psi$.
The limits include  
the constraints on millicharged particles (shaded light gray) 
\cite{Davidson:2000hf, Acciarri:2019jly, Ball:2020dnx},
the electroweak precision measurements for
the $Z$ mass shift (dashed red) \cite{Du:2019mlc}, 
the $Z$ invisible decay (dashed green)~\cite{ALEPH:2005ab}, 
the dilepton high mass resonance search 
at ATLAS ({dash-dotted} blue) ~\cite{ATLAS:2019vcr}, 
and the {monojet} search at ATLAS (solid black) 
\cite{Aaboud:2017phn}.
}
\label{fig:limit-mchi-eps2} 
\end{centering}
\end{figure}

Fig.~\ref{fig:limit-mchi-eps2} shows various experimental 
constraints on the model, including 
the 
{constraints}
from millicharged particle searches
\cite{Davidson:2000hf, Acciarri:2019jly, Ball:2020dnx},
the electroweak precision measurements for
the $Z$ mass shift \cite{Du:2019mlc}, 
the $Z$ invisible decay \cite{ALEPH:2005ab}, 
the dilepton high mass resonance search 
at ATLAS \cite{ATLAS:2019vcr}, 
and the {monojet} search at ATLAS 
\cite{Aaboud:2017phn}. 
Here we choose $m_1 = 3$ GeV, 
$m_2 = 700$ GeV, $g_F = 1.5$, $g_W = 1$, 
and $\epsilon_1 \ll \epsilon_2$. 
Throughout this analysis 
we use $m_2 = 700$ GeV, $g_F = 1.5$, and $g_W = 1$ 
as the default values for these three parameters 
as in {Ref.\ \cite{Du:2019mlc}};
in the {parameter} space
of interest, we have 
$m_1 \simeq$ GeV $\ll m_2$ 
so that the dark photon mass 
$m_{A'} \simeq m_1$, 
and the heavy $Z'$ boson 
has a mass $m_{Z'} \simeq m_2$.
For {the hidden fermion mass} $m_\psi  \gtrsim 3$ GeV 
the electroweak constraint on the $Z$ mass shift 
gives the most stringent limit, 
$\epsilon_2 \lesssim 0.036$, 
whereas for the mass range 
$0.3$ GeV $\lesssim m_\psi  \lesssim 3$ GeV, 
the leading constraints come from 
the recent ArgoNeuT data \cite{Acciarri:2019jly}
and the milliQan demonstrator data \cite{Ball:2020dnx}. 
We note that the mass fraction of the millicharged DM 
is constrained to be $ \lesssim 0.4\%$ by the CMB data 
\cite{Boddy:2018wzy, dePutter:2018xte, Kovetz:2018zan}, 
which is satisfied in the parameter 
space of interest of our model \cite{Du:2019mlc}.

\section{A mini-overview on lifetime-frontier detectors}
\label{sec: detector-review}

A number of new lifetime-frontier 
detectors have been proposed recently 
at the LHC, 
which can be used to search for LLPs. 
Table \ref{tab:detectors} shows 
the angular coverage,
location, size, and expected running time 
of these new detectors.    
We classify the detectors into three 
categories: forward detectors, 
central transverse detectors, 
and precision timing detectors. 
The forward detectors include 
FACET \cite{FACET:talk1, Cerci:2021nlb}, 
FASER \cite{Feng:2017uoz, Ariga:2018zuc, 
Ariga:2018uku, Ariga:2018pin, Ariga:2019ufm}, 
{FASER2} \cite{ Ariga:2019ufm, Kling:2021fwx}, 
AL3X \cite{Gligorov:2018vkc}, 
and {MoEDAL-MAPP} \cite{Staelens:2019gzt}. 
The central transverse detectors include 
CODEX-b \cite{Gligorov:2017nwh, Aielli:2019ivi}, 
MATHUSLA 
\cite{Chou:2016lxi, Curtin:2018mvb, Alpigiani:2018fgd, 
Lubatti:2019vkf, Alpigiani:2020tva}, 
ANUBIS \cite{Bauer:2019vqk}. 
The precision timing detectors include 
CMS-MTD \cite{CMStiming}, 
ATLAS-HGTD \cite{Allaire:2018bof}, 
and LHCb-TORCH~\cite{LHCb:2017MTD, LHCb:2008vvz}. 
Below we provide a mini-overview 
of the new lifetime frontier detectors.

\begin{table*}[htbp]
\begin{tabular}{|l|c|c|c|c|c|}  
\hline
 Detector
&  \multicolumn{1}{|p{1.5cm}|}{\centering $\eta$}
& \multicolumn{1}{|p{4cm}|}{\centering Distance from IP (m)}
& \multicolumn{1}{|p{3cm}|}{\centering Decay volume (m$^3$)}
& \multicolumn{1}{|p{2cm}|}{\centering LHC runs}
\\ \hline 
\hline
FACET~\cite{FACET:talk1, Cerci:2021nlb}
& $[6, 7.2]$
& $100$  ({upstream}) 
& $12.3$
& run 4 (2027)
\\ 
\hline 
FASER~\cite{Feng:2017uoz, Ariga:2018zuc, 
Ariga:2018uku, Ariga:2018pin, Ariga:2019ufm}
& $> 9$
& $480$ (downstream) 
& $0.047$
& run 3 (2022) 
\\ 
\hline 
{FASER2} \cite{ Ariga:2019ufm, Kling:2021fwx}
& $> 6.87$
& $480$ (downstream) 
& $15.7$
& HL-LHC 
\\ 
\hline
AL3X~\cite{Gligorov:2018vkc}
& $[0.9, 3.7]$ 
& $5.25$ ({upstream}) 
& $915.2$ 
& run 5 ({2032})
\\ 
\hline
{MoEDAL-MAPP} \cite{Staelens:2019gzt}
& $\sim 3.1$
& $55$  ({upstream}) 
& $\sim 150$ 
& run 3 (2022)
\\   
\hline
\hline
CODEX-b~\cite{Gligorov:2017nwh, Aielli:2019ivi}
& $[0.14, 0.55]$ 
& $26$  (transverse) 
& $10^3$ 
& run 4 (2027)
\\ 
\hline
MATHUSLA 
\cite{Chou:2016lxi, Curtin:2018mvb, Alpigiani:2018fgd, 
Lubatti:2019vkf, Alpigiani:2020tva} 
& $[0.64, 1.43]$ 
& $60$ (transverse) 
& $2.5 \times 10^5$ 
& HL-LHC
\\ 
\hline
ANUBIS~\cite{Bauer:2019vqk}
& [0.06, 0.21]
& $24$  (transverse) 
& $\sim 1.3 \times 10^4$ 
& HL-LHC
\\   
\hline
\hline
CMS-MTD~\cite{CMStiming}
& $[-3, 3]$
& {$1.17$ (barrel), $3.04$ (endcaps)}
& 25.4 
& HL-LHC 
\\   
\hline
ATLAS-HGTD~\cite{Allaire:2018bof}
& $[2.4, 4]$ 
& $ 3.5$  (endcaps)
& $8.7$
& HL-LHC 
\\   
\hline
LHCb-TORCH~\cite{LHCb:2017MTD, LHCb:2008vvz}
&  $[1.6, 4.9]$ 
& $9.5 $ {(beam direction)} 
& --
& HL-LHC 
\\ 
\hline
\end{tabular}  
\caption{Proposed detectors for long-lived particles searches at the LHC.
The first column shows the detector name,
the second column shows the pseudorapidity coverage,
the third column shows the distance from {interaction point (IP)}  
to the near side of the detector 
and the location 
(to the far side of the detector, for FASER), 
the fourth column shows the decay volume of the detector, 
and the last column shows the starting time of {data-taking}. 
The first {five} detectors are located at the 
forward region of the {corresponding} IP; 
the middle three detectors are located 
at the far central {transverse} region of the 
{corresponding} IP; 
the last three detectors are the 
precision timing detectors 
to be installed at CMS, ATLAS and LHCb 
respectively to control the HL-LHC 
pile-up {background}. 
The HL-LHC is expected to start 
{data-taking} in 2027 (run 4) \cite{LHCtime}. 
Here ``upstream'' (``downstream'')
means that the detector is located in
the clockwise (anticlockwise) direction
of the {corresponding} IP, 
{viewed from above}.}
\label{tab:detectors}
\end{table*}

\subsection{Forward detectors}

FASER 
(the ForwArd Search ExpeRiment), 
is located at $480$ m downstream of the ATLAS detector 
along the beam axis 
\cite{Feng:2017uoz, Ariga:2018zuc, Ariga:2018uku, Ariga:2018pin, Ariga:2019ufm}. 
FASER has a cylindrical 
decay volume {of length} ${L}  = 1.5$ m and  radius $R = 10$ cm. 
FASER has been installed at {the} TI12 tunnel at {the} LHC 
and 
{is expected} to collect data during LHC Run 3 (2022) 
\cite{FASER:2019dxq}.
The upgrade version, FASER 2, with {a decay volume of length} ${L = 5}$ m
and radius $R = 1$ m 
is proposed to be installed during the HL-LHC run (2026-35) 
\cite{ Ariga:2019ufm, Kling:2021fwx}.

\begin{figure}[htbp]
\begin{centering}
\includegraphics[width=0.4\textwidth]{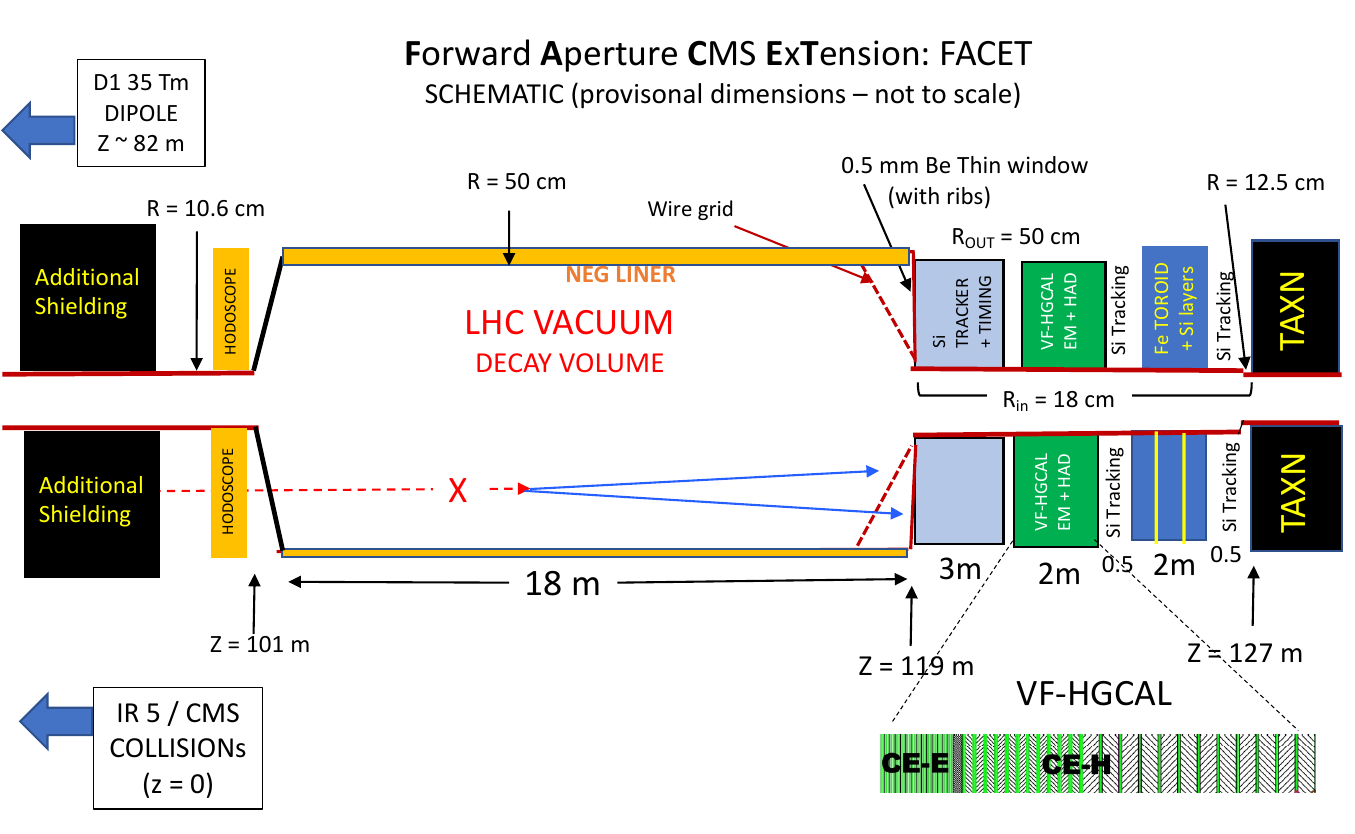}
\caption{Schematic layout of the proposed 
FACET detector (side view) \cite{albrow}.}
\label{fig:facet-layout1} 
\end{centering}
\end{figure}

FACET 
(Forward-Aperture CMS ExTension) 
is a new lifetime frontier detector which 
is proposed to be installed $\sim$100 m
upstream of the CMS detector 
along the beam axis \cite{FACET:talk1, Cerci:2021nlb}.
FACET {is proposed to} be built based on the CMS Phase 2 Upgrade concept, 
combining silicon tracker, timing detector, HGCAL-type EM/HAD calorimeter, 
and GEM-type muon system in a compact design \cite{FACET:talk1, Cerci:2021nlb, FACET:talk2}; 
the latest design of the  
FACET detector is shown in Fig.\ \ref{fig:facet-layout1}.  
The decay volume of the FACET experiment 
is an enlarged LHC quality vacuum beam pipe 
which is 18 m long and has a radius of 50 cm
\cite{FACET:LOI, FACET:talk1, Cerci:2021nlb, FACET:talk2}. 
The FACET detector is shielded by about {35}-50 m
of steel (in the Q1-Q3 quadrupoles and D1 dipole)
in front of it \cite{FACET:LOI}; 
additional shielding materials are placed before the decay volume, 
as shown in Fig.\ \ref{fig:facet-layout1}. 
The FACET detector, 
surrounding the LHC beam pipe which has a 
radius $R = 18$ cm, is placed behind  
the decay volume. 
As a new proposed far forward detector, FACET has some {merits}. 
The {35}-50 m steel shielding before FACET, 
corresponding to $200-300$ nuclear interaction lengths, 
is comparable to the shielding material for FASER, which 
is $\sim 100$ m of concrete/rock, 
corresponding to $\sim 240$ nuclear interaction lengths. 
FACET will benefit from the high quality LHC vacuum 
pipe as the decay volume \cite{FACET:talk1, Cerci:2021nlb, FACET:talk2}. 
FACET plans to have both the EM 
and HAD calorimeters \cite{FACET:LOI}, 
whereas FASER has only EM calorimeter
\cite{Ariga:2018zuc, Ariga:2018uku, Ariga:2018pin, Ariga:2019ufm}. 
This allows FACET to have a better detection efficiency  
for the hadronic decays of the DP, 
especially for the neutral hadronic decays.

AL3X
(A Laboratory for Long-Lived eXotics) 
is {an on-axis cylindrical detector} which
has been proposed to be installed 
at ALICE experiment during the LHC Run 5 \cite{Gligorov:2018vkc}.  
The detector will make use of 
the existing ALICE time projection chamber and the L3 electromagnet. 
It is also envisioned to move the ALICE detector by 11.25 m downstream from its current location, 
providing space for a spherical shell segment of tungsten to shield the detector from the IP. 
The AL3X detector is then expected to be located $5.25$ m away from the IP along the beam axis,  
with a 12 m long cylindrical decay volume of a 0.85 m inner radius 
and a 5 m outer radius.

The MoEDAL-MAPP {detector}
is the MAPP 
(Apparatus for the detection of Penetrating Particles) 
detector at MoEDAL 
(Monopole and Exotics Detector at the LHC) 
\cite{Staelens:2019gzt}, which 
is proposed to be installed at the UGCI gallery
near the LHCb experiment (IP8) in future LHC runs. 
MoEDAL-MAPP is 
$55$ m from IP8 
and with an angle of $5^{\circ}$ away 
from the beam line, 
with a fiducial volume 
of $\sim$150 m$^3$ 
\cite{Staelens:2019gzt}.

\subsection{Central detectors}

CODEX-b
(Compact Detector for Exotics at LHCb)  
has been proposed to be constructed at the LHCb 
cavern
\cite{Gligorov:2017nwh, Aielli:2019ivi}. 
The decay volume is designed to be  
$10\, \rm{m} \times 10 \,\rm{m} \times 10\, \rm{m}$.
It is located $\sim$5 m in the z axis (beam direction) 
and $\sim$26 m in the transverse direction away 
from the LHCb IP, 
with a pseudorapidity coverage of $0.14 <\eta< 0.55$.
The demonstrator detector, CODEX-$\beta$ 
(about $2\, \rm{m} \times 2 \,\rm{m} \times 2\, \rm{m}$)
has been developed for the LHC Run 3 \cite{Aielli:2019ivi}.

MATHUSLA
(MAssive Timing Hodoscope for Ultra-Stable neutraL pArticles) 
is a new proposed experiment 
near the ATLAS or CMS {IP}  
\cite{Chou:2016lxi, Curtin:2018mvb, Alpigiani:2018fgd, 
Lubatti:2019vkf, Alpigiani:2020tva}. 
It is proposed to {be} placed $\sim$68 m downstream from the IP 
and $\sim$60 m above the LHC beam axis
with a decay volume  
of $100\,  \rm{m} \times 100\,  \rm{m} \times 25\,  \rm{m}$ 
\cite{Alpigiani:2020tva}. 
MATHUSLA was previously  
proposed to be installed at
$\sim$100 m downstream from the IP 
and $\sim$100 m above the LHC beam axis 
with a decay volume of  
$200\,  \rm{m} \times 200\,  \rm{m} \times 20\,  \rm{m}$ 
\cite{Chou:2016lxi, Curtin:2018mvb, Alpigiani:2018fgd, Lubatti:2019vkf}. 
In this analysis, we adopt the parameters from the  
recent proposal \cite{Alpigiani:2020tva}.

ANUBIS
(AN Underground Belayed In-Shaft search experiment) 
\cite{Bauer:2019vqk} 
is a {new proposed} experiment taking advantage of 
the 18 m diameter, 
56 m long PX14 installation shaft 
of the ATLAS experiment. 
The proposed detector consists of four tracking stations 
which have the same cross section area of 230 m$^2$
and are 18.5 m apart from each other.

\subsection{Precision timing detectors}

To mitigate the high pile-up background at the
HL-LHC, various precision timing detectors will 
be installed at
CMS \cite{CMStiming},
ATLAS \cite{Allaire:2018bof, Allaire:2019ioj,Garcia:2020wxj},
and LHCb \cite{LHCb:2017MTD}, 
which can be used for LLP searches 
\cite{Liu:2018wte, Mason:2019okp, Kang:2019ukr,
Cerri:2018rkm, Du:2019mlc, Liu:2020vur, 
Bhattacherjee:2020nno, Cheung:2021utb}.\footnote{{For early studies 
on precise timing, see e.g., Ref.~\cite{Shrock:1978ft}, 
where the 
precise timing method was proposed to search for LLPs 
or stable particles in a beam dump experiment. 
Such a search was later carried out in the E733 experiment 
at Fermilab \cite{FMMF:1994yvb}}.}

The CMS-MTD {detector}
consists of 
the precision minimum ionizing particle (MIP) timing detector
with a timing resolution of 30 picoseconds 
\cite{CMStiming}. 
The timing layers will be installed between  
the inner trackers and the electromagnetic calorimeter 
for the barrel and {endcap} regions. 
The timing detector in the barrel region
has a length of 6.08 m along the beam axis direction
and a transverse distance of $1.17$ m away from the beam.
The timing detectors in the {endcap} regions
have a pseudorapidity coverage of $1.6 < | \eta | < 3.0$
and are located $\sim 3.0$ m from the IP. 
The decay volume of LLPs at CMS-MTD is 
$\sim 25.4$ ${\rm m}^3$ if one demands that 
the LLPs decay before arriving {at} the timing layers and 
the decay vertex has a transverse distance of  
$0.2\, {\rm m} < L_T < 1.17\, {\rm m}$ from the beam 
axis \cite{Liu:2018wte, Liu:2020vur}.

The HGTD (High Granularity Timing Detector) 
has been proposed to be installed
in front of the ATLAS endcap and forward calorimeters
at $z = \pm 3.5$ m from the IP
during the ATLAS Phase-II upgrade
\cite{Allaire:2018bof, Allaire:2019ioj, Garcia:2020wxj}.
The ATLAS-HGTD can cover the pseudorapidity range
of $2.4 < | \eta | < 4.0$, and
is expected to have a time resolution of 35 ps (70 ps) per hit
at the start (end) of HL-LHC \cite{Garcia:2020wxj}.
The decay volume of ATLAS-HGTD is $\sim 8.7 \,\rm{m}^3$, 
if LLPs are required to decay before arriving {at} the timing detector and 
the decay vertex has a transverse distance of  
$0.12\, {\rm m} < L_T < 0.64\, {\rm m}$ 
\cite{Garcia:2020wxj}.

The TORCH (Time Of {internally} Reflected CHerenkov light) detector
has been proposed to be installed at the next upgrade of LHCb \cite{LHCb:2017MTD}. 
The TORCH will be located at $z\sim 9.5$ m from the LHCb IP
with {the} angular acceptance of 1.6$<\eta<$4.9. 
The precision of each track in the TORCH system is 15 ps \cite{LHCb:2017MTD}.

\section{The dark photon production}
\label{sec:DP_production}

\begin{figure*}[htbp]
\begin{centering}
\includegraphics[width=0.28\textwidth]{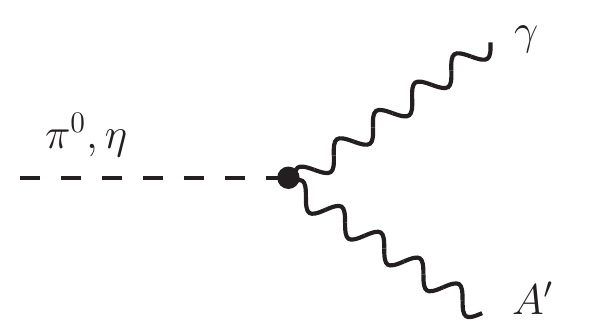}
\includegraphics[width=0.3\textwidth]{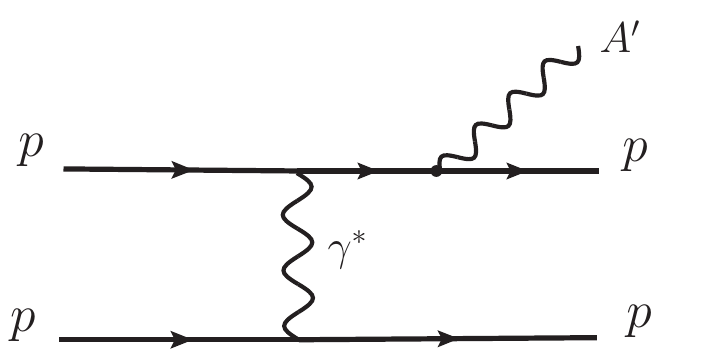}
\includegraphics[width=0.26\textwidth]{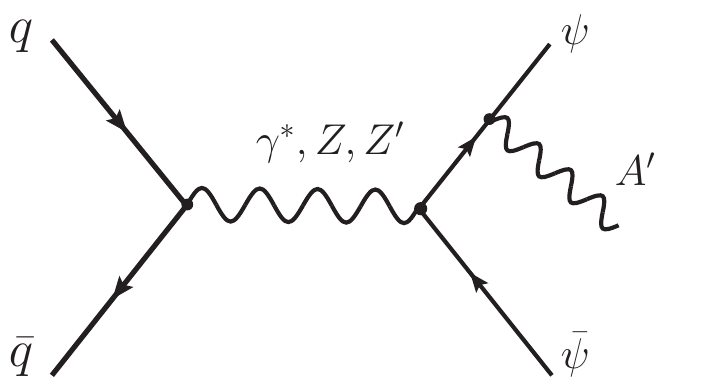}
\caption{
Feynman diagrams for the dark photon production at the LHC: 
from meson decays (left), 
from the proton bremsstrahlung (middle), 
and from the hidden fermion radiation (right).
}
\label{fig:feyndiag} 
\end{centering}
\end{figure*}

In our model, there are three main processes to produce 
dark photon $A'$ at the LHC: 
rare meson decays (hereafter MD), 
coherent proton bremsstrahlung (hereafter PB), 
and hidden sector radiation (hereafter HR); 
the corresponding Feynman diagrams are shown 
in Fig.\ (\ref{fig:feyndiag}). 
The MD and PB processes are common 
for the dark photon models, 
because dark photons are produced via  
interactions between the dark photon and 
charged particles in the SM 
in these two processes. 
The HR process is new in our model \cite{Du:2019mlc}, 
which is mediated by the interaction between   
the dark photon and the hidden sector particle $\psi$.\footnote{Here 
we do not consider 
the dark photon direct production channel which 
consists of the following processes 
$q\bar{q} \to A'$, $q\bar q \to g A'$, $q g\to q A'$ and $\bar{q} g \to  \bar{q} A'$,
because they suffer from large PDF uncertainties  
for sub-GeV $A'$
and are suppressed by $\epsilon_1$ 
which is much smaller than $\epsilon_2$ in the HR process.}

\subsection{Meson decays}
\label{subsec:MD}

Dark photons can be produced in 
the $m \to \gamma + A'$ process, 
where $m$ denotes a light meson, 
as shown in the left diagram in 
Fig.\ \ref{fig:feyndiag}; 
the branching ratio can be computed via \cite{Batell:2009di}
\be
{\rm BR} \left (m \to A' + \gamma \right) = 2\, \epsilon^{2} 
\left( 1-\frac{M_{A'}^2}{M_{m}^2} \right)^{3} 
{\rm BR}\left(m \to \gamma \gamma \right), 
\label{eq:brm}
\ee
{where $\epsilon$ is the coupling constant given in Eq.\ \eqref{eq:epsilon}.}
In the parameter space of interest of our model, 
one has $\epsilon \approx (0.27/e)\, \epsilon_1$ for $m_1 \lesssim 30$ GeV. 
For light mesons, one has
${\rm BR}(\pi^{0} \rightarrow \gamma \gamma ) \simeq 0.99$ 
and ${\rm BR}(\eta \rightarrow \gamma \gamma) \simeq 0.39$ \cite{Zyla:2020zbs}.
Since light mesons can be copiously produced 
in the forward direction at  
high energy $pp$ collisions,
(for example, the 
production cross section of $\pi^0$ ($\eta$) 
in each hemisphere at the LHC  
is $1.6 \times 10^{12}$ pb 
($1.7 \times 10^{11}$ pb) \cite{Ariga:2018uku}),  
dark photon from rare meson decays 
can be a leading dark photon production mode at the LHC 
if the decay is kinematically allowed 
\cite{Feng:2017uoz}. 
We neglect the $m \to A' A'$  process 
because we have $\epsilon \ll 1$ 
for the LLDP.

In our analysis, 
we generate the {four-momentum spectrum} 
for the $\pi^0/\eta$ mesons 
using the EPOS-LHC \cite{Pierog:2015} model in  
CRMC \cite{Ulrich:crmc} 
with $10^5$ simulation events of $pp$ inelastic collision at the 
LHC with $\sqrt{s}=13$ TeV. 
We then boost the momentum of the dark photon 
(which is isotropically distributed in the $\pi^0/\eta$ rest frame) 
to the lab frame, by using the meson momentum.
Our simulations are found to be consistent with 
FORESEE \cite{Kling:2021fwx}. 
{We also simulate the heavy mesons $D^0$, $B^0$ and $J/\psi$ 
using PYTHIA 8 \cite{Sjostrand:2014zea}. 
We found that the DP production cross section 
due to decays of these heavy mesons 
is about five orders of magnitude smaller than the light mesons ($\pi^0$ and $\eta$).
Therefore we neglect the contribution from heavy meson decays in our analysis.}

\subsection{Proton bremsstrahlung}
\label{subsection:proton_bremss}

Proton bremsstrahlung process 
is another major production mode of 
light dark photons in high energy 
$pp$ collisions; the Feynman 
diagram  
is shown as the middle diagram 
in Fig.\ \ref{fig:feyndiag}.
The dark photon signal arising from 
the proton bremsstrahlung process 
can be computed by the 
Fermi-Weizsacker-Williams (FWW) method 
\cite{Fermi:1924, Williams:1934, Weizsacker:1934}, 
in which {the} proton is treated as a coherent object; 
the total number of the dark photon produced 
in a far forward detector\footnote{For a near 
detector with nearly $4\pi$ coverage, 
e.g., CMS, one can use the FWW method to compute the  
PB contributions from each colliding proton in the 
lab frame.}
is given by \cite{Feng:2017uoz}
\bea
N_{A'}^{\rm PB} &=& %
{\cal L} \, 
{|F_{1}\left(m_{A^{\prime}}^{2}\right)|^{2}}
\int d z \, d p_{T}^{2} \,\sigma_{p p}\left(s^{\prime}\right)
w\left(z, p_{T}^{2}\right) \nonumber \\
&& \times \Theta\left(\Lambda_{\mathrm{QCD}}^{2}-q_{\rm{min}}^{2}\right),  
\label{eq:proton-brem}
\eea
where $N_{A'}^{\rm PB}$ is the number of dark photon
events from the PB process,
${\cal L}$ is the integrated luminosity,
$F_1$ is the form factor function, 
$z=p^L_{A'}/p_p$ with $p^L_{A'}$ being 
the longitudinal momentum of 
the dark photon  
and $p_p$ the proton beam momentum,
$p_T$ is the transverse momentum of the dark photon, 
$\sigma_{pp}(s^{\prime})$ is the inelastic cross section 
\cite{Tanabashi:2018oca} with $s' = 2 m_p (E_p -E_{A'})$ 
{in the rest frame of one of the colliding protons}, 
$w\left(z, p_{T}^{2}\right)$ is the splitting function,  
$\Lambda_{\mathrm{QCD}} \simeq 0.25$ GeV 
is the QCD scale, 
and $q$ is the momentum carried by the virtual photon 
in the middle diagram 
in Fig. \ref{fig:feyndiag}. 
The splitting function $w\left(z, p_{T}^{2}\right)$ in Eq.~\eqref{eq:proton-brem} 
is given by~\cite{Brunner:2014, Kim:1973, Tsai:1977}
\begin{widetext}
\bea
w\left(z, p_{T}^{2}\right)& \simeq & \frac{\epsilon^{2} \alpha}{2 \pi H}\left\{\frac{1+(1-z)^{2}}{z}-2 z(1-z)\right.
 \left(\frac{2 m_{p}^{2}+m_{A^{\prime}}^{2}}{H}-z^{2} \frac{2 m_{p}^{4}}{H^{2}}\right) \\
&& + \,\; 2 z(1-z)\left(z+(1-z)^{2}\right) \frac{m_{p}^{2} m_{A^{\prime}}^{2}}{H^{2}} \left. + 2 z(1-z)^{2} \frac{m_{A^{\prime}}^{4}}{H^{2}}\right\}, 
\eea
\end{widetext}
where $H=p_{T}^{2}+(1-z) m_{A^{\prime}}^{2}+z^{2} m_{p}^{2}$. 
To guarantee the validity of the FWW  
approximation, the Heaviside function $\Theta$ is imposed in Eq.~\eqref{eq:proton-brem} 
with the minimal virtuality of the photon 
cloud around the beam proton given by \cite{Kim:1973, Tsai:1977} 
\be
\left|q_{\min }^{2}\right| \approx \frac{1}{4 E_{p}^{2} z^{2}(1-z)^{2}}\left[p_{T}^{2}+(1-z) m_{A^{\prime}}^{2}+z^{2} m_{p}^{2}\right]^{2}. 
\ee
The form factor $F_1(p_{A'}^2)$ in Eq.~\eqref{eq:proton-brem} is given by~\cite{Feng:2017uoz, Faessler:2009tn}
\be
F_1(p_{A'}^2) = \sum_{V =\rho \, \rho' \, \rho'' \, \omega \, \omega' \, \omega''} \frac{f_V m_V^2}{m_V^2 - p_A'^2 - i m_V \Gamma_V}, 
\label{eq:PB_formfactor}
\ee
where $m_V$ ($\Gamma_V$) is the mass (decay width) of the vector meson, 
$f_{\rho} = 0.616$, $f_{\rho'} = 0.223$, $f_{\rho''} = -0.339$, $f_{\omega} = 1.011$, $f_{\omega'} = -0.881$, and $f_{\omega''} = 0.369$.

\subsection{Hidden radiation}

Dark photons can also be produced via hidden fermion 
radiations in the HR process, 
as shown in the third diagram of Fig.\ \ref{fig:feyndiag}.
Within certain parameter 
space of the models in {Ref.\ \cite{Du:2019mlc}},
the HR process can be more important 
than the MD and PB processes.  
For the models considered in this analysis,  
dark photons in the HR process 
are produced at the LHC 
in the radiation process of 
the hidden sector fermion $\psi$, 
which are pair-produced at the 
LHC via the $q \bar q \to \gamma^{*}/Z/Z' \to \bar \psi \psi$ 
process, 
as shown in the third diagram in Fig.~\ref{fig:feyndiag}.

In the MD and PB processes, 
the dark photon production cross section 
is suppressed by the small $\epsilon$ parameter 
(given in Eq.\ \eqref{eq:epsilon})
needed for the long lifetime of the dark photon. 
In the HR process, however, 
the LHC production of $\psi$ is not 
controlled by $\epsilon$ so that 
the LHC cross section {of} $\psi$
can be {sizable} even for heavy 
$\psi$.

\begin{figure}[htbp]
\includegraphics[width=0.4\textwidth]{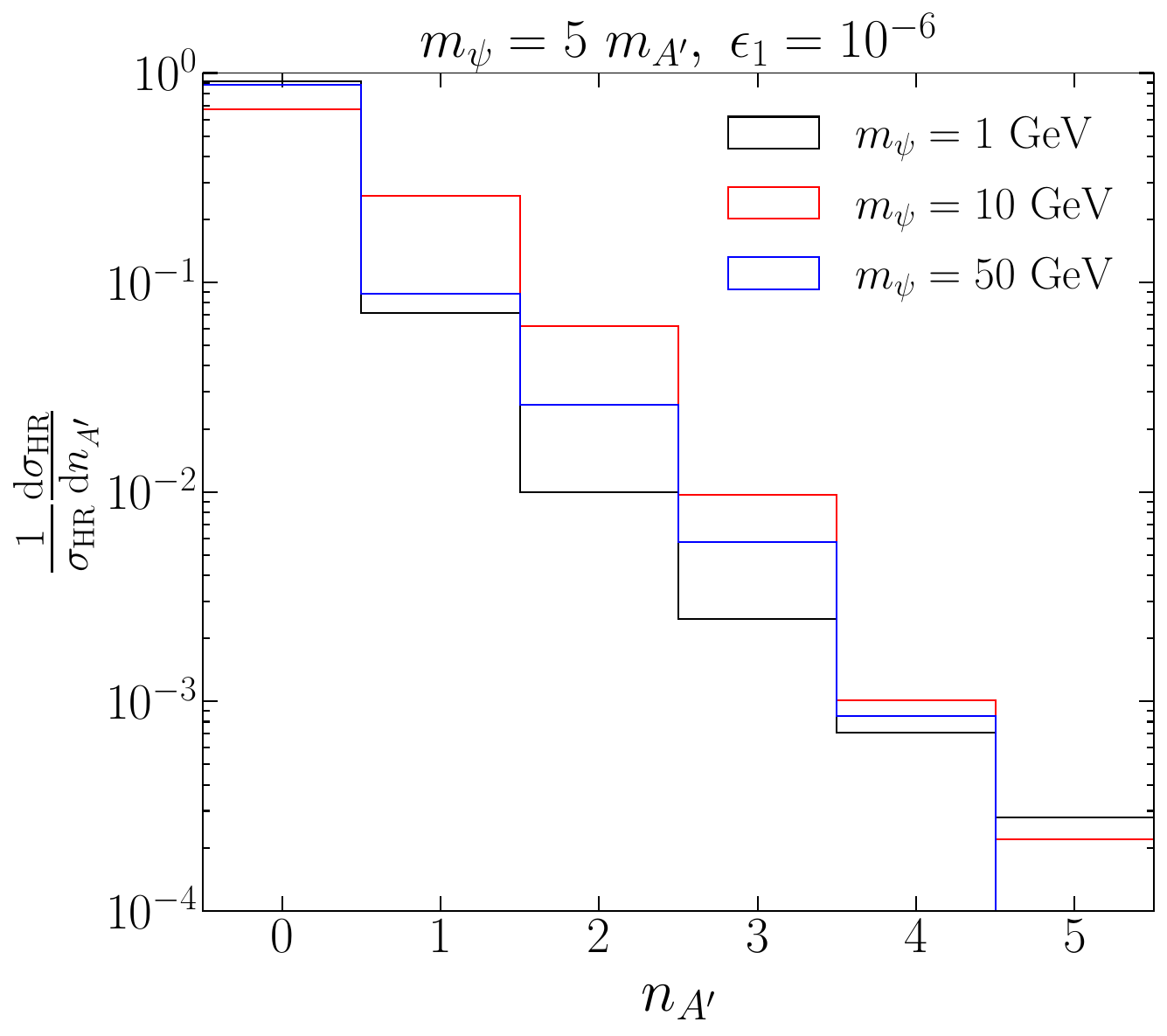}
\caption{The normalized distribution of 
dark photon multiplicity in the $\psi \bar\psi$ events, 
where $\sigma_{\rm HR}$ is the  $\psi \bar\psi$ cross section. 
We take $m_{\psi} = 5\ m_{A'}$ and 
$\epsilon_1=10^{-6}$. 
The black, red, and blue histograms  
are for the $m_\psi = $ 1 GeV, 10 GeV, 
and 50 GeV cases respectively.} 
\label{fig:n_dp_dis}
\end{figure}

\begin{figure}[htbp]
\includegraphics[width=0.4\textwidth]{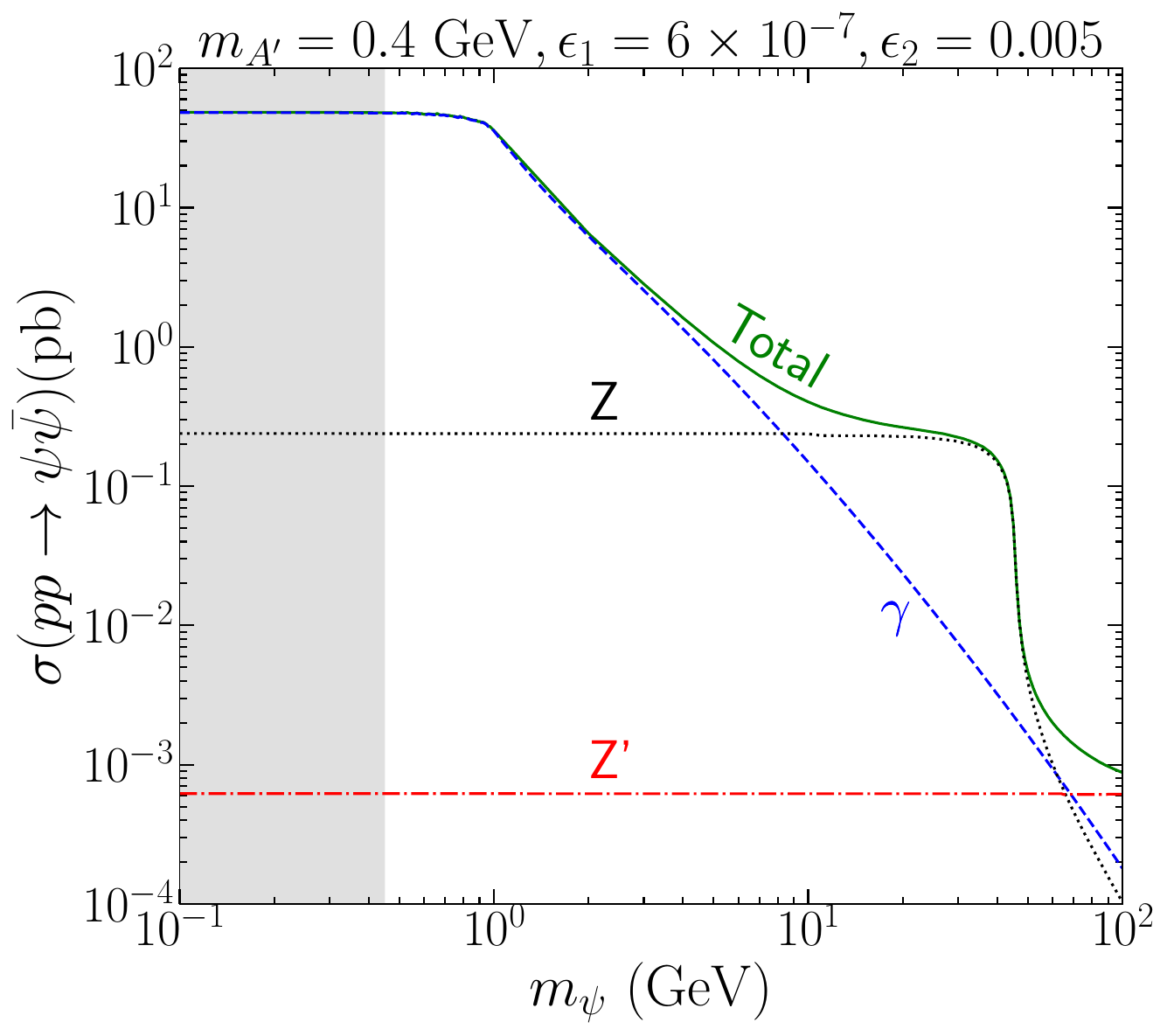}
\caption{The contributions 
to the $pp \rightarrow  \psi\bar{\psi}$ cross section 
at the LHC 
from three different mediators:  
$\gamma$ (blue-dashed), 
$Z$ (black-dotted), 
$Z'$ (red-{{dashdotted}}). 
{The total cross section (green-solid) 
taking into account all contributions 
(including the $A'$ contribution and the 
interference terms) is also shown.}  
We use 
$\epsilon_{1}=6\times 10^{-7}$, 
$\epsilon_{2}= {0.005}$,
and $m_{A'} = 0.4$ GeV. 
The gray shaded region 
indicates the parameter space 
excluded by the millicharge constraints
\cite{Davidson:2000hf, Acciarri:2019jly, Ball:2020dnx}. 
We use {NNPDF23LO} \cite{Ball:2012cx} which 
is the default PDF in MADGRAPH 5.}
\label{fig:prop-xsec-psipair}
\end{figure}

To obtain the contribution from the HR process, 
we use FeynRules \cite{Alloul:2013bka} 
to produce the UFO file for our model, 
which is then passed into MADGRAPH 5 \cite{Alwall:2014hca} 
to generate the $p p \to \psi \bar{\psi}$ events at the LHC. 
We further use PYTHIA 8 
\cite{Sjostrand:2014zea, Carloni:2010tw, Carloni:2011kk} 
to simulate the dark radiation process of the $\psi$ particle 
to obtain the dark photons.

The dark photon cross section 
in the HR process at the LHC can be computed via  
\be
\sigma_{A'}^{\rm HR} =  \bar{n}_{A'} \, \sigma(p p \to \psi \bar\psi), 
\ee
where $\bar{n}_{A'}$ is the expected number 
of dark photons per $\psi \bar \psi$ event, 
and $\sigma(p p \to \psi \bar\psi)$ is the 
production cross section of the $\psi \bar\psi$ 
events at the LHC. 
We compute $\bar{n}_{A'}$ by taking    
the ratio of the total number of dark photons 
in our PYTHIA simulation 
to the number of $\psi \bar \psi$ events simulated.
We note that multiple dark photons can be 
radiated by one $\psi$ fermion; 
see e.g., refs.\ 
\cite{Bai:2009it, Buschmann:2015awa, Kim:2016fdv, Chen:2018uii} 
for some earlier studies on dark vector bosons radiated from  
hidden sector fermions. 
Fig.\ (\ref{fig:n_dp_dis}) shows the normalized distribution
of the number of dark photons in 
the $\psi \bar \psi$ events 
for three benchmark models in our simulation with the relation 
$m_\psi = 5 m_{A'}$; 
the expected dark photon number are  
$\bar{n}_{A'} \simeq $ 0.097, 0.42 and 0.16
for $m_{\psi} = $ 1, 10 and 50 GeV cases respectively.

{We compare three different contributions 
to $\sigma(p p \to \psi \bar\psi)$ at the LHC
from three different mediators 
(photon, $Z$, and $Z'$) in Fig.~\ref{fig:prop-xsec-psipair},}
where the interference effects have been 
neglected.\footnote{We neglect the process mediated 
by the dark photon since it is suppressed by the 
small $\epsilon$ parameter needed for LLDP so that 
it is several orders of magnitude smaller than 
the other three mediators in our analysis. 
}
We use MADGRAPH 5 \cite{Alwall:2014hca} 
to compute the cross sections, where 
we have fixed 
$m_{A'} = 0.4$ GeV, 
$\epsilon_1 = 6 \times 10^{-7}$, 
and $\epsilon_2 = {0.005}$.
For $m_{\psi} \lesssim 8$ GeV
the dominant contribution to the 
$\psi \bar \psi$ pair-production cross section comes  
from the s-channel photon process;
for higher $\psi$ mass, 
the contributions from $Z$ and $Z'$ 
exchanges become more important.

\subsection{Comparison of the three DP production channels}

\begin{figure*}[htbp]
\includegraphics[width=0.4\textwidth]{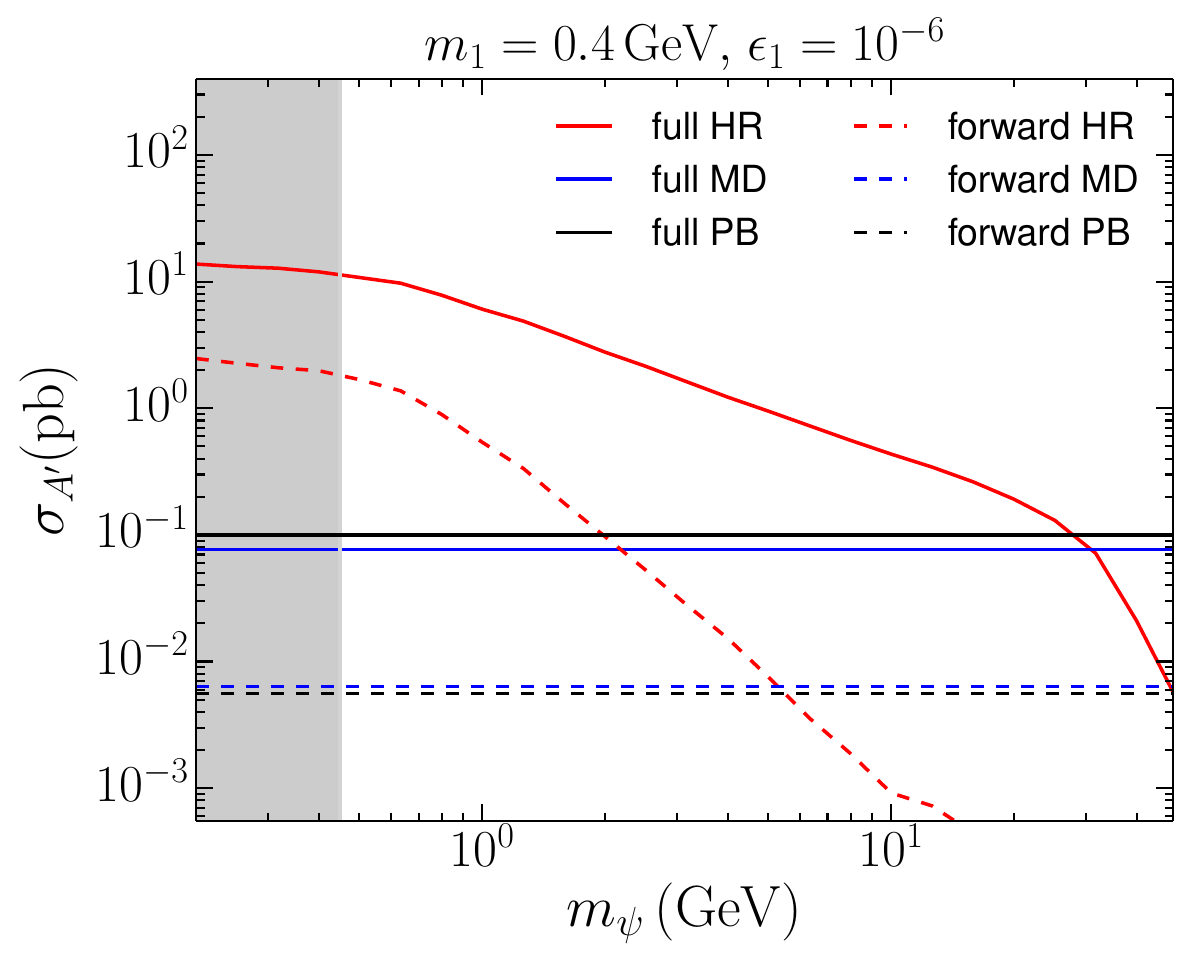}
\includegraphics[width=0.4\textwidth]{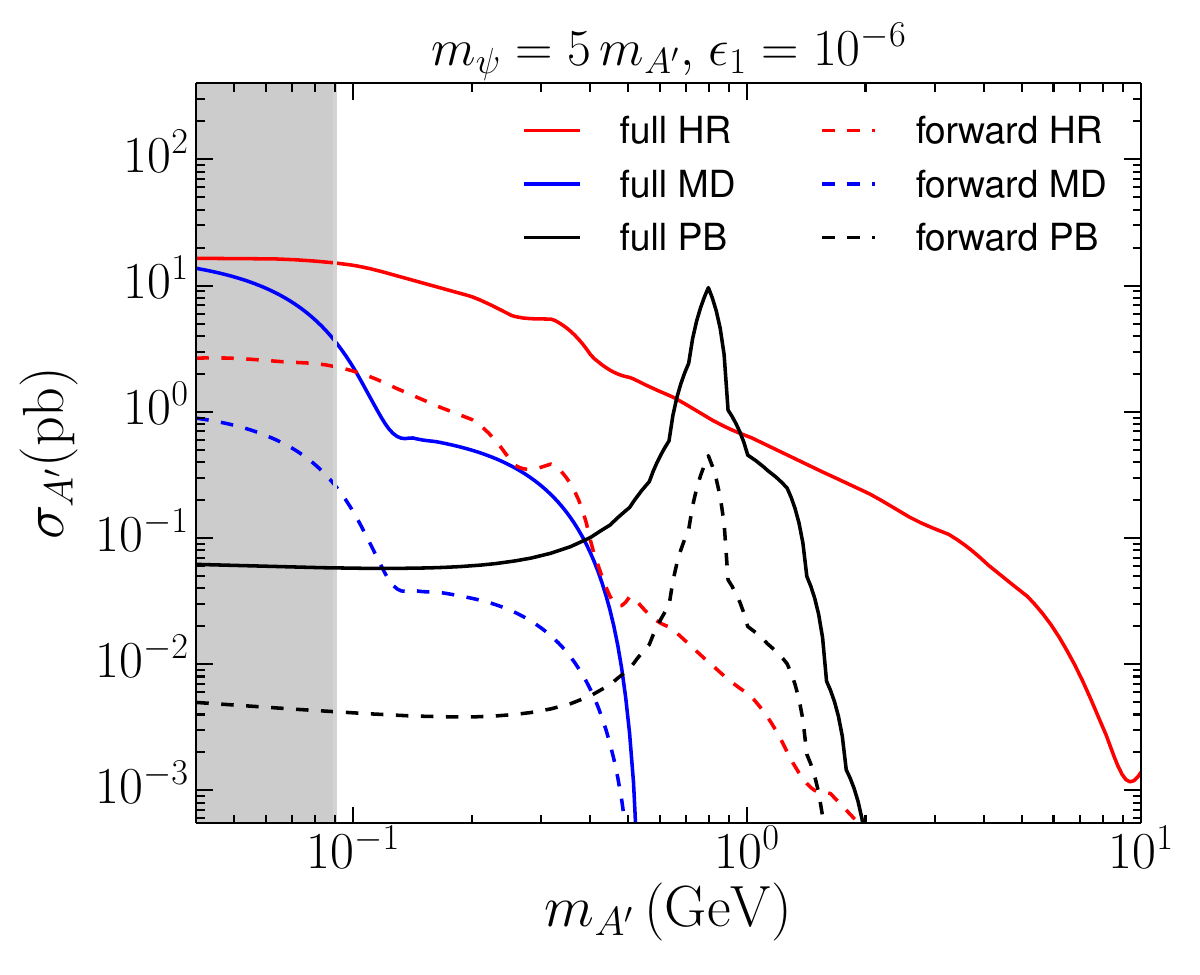}
\caption{The LHC production cross section of dark photon $\sigma_{A'}$ 
from three contributions: HR (red lines), MD (blue lines) and PB (black lines). 
The solid lines correspond to the cross section in full solid angle, 
and the dashed lines represent the cross section in 
the very forward region with $\eta_{A'} > 6$. 
Here we use $\epsilon_1 = 10^{-6}$ and $\epsilon_2 = 0.005$ for both panels;  
we choose $m_{1} = 0.4$ GeV in the left panel 
and $m_{1} = 0.2 \,m_{\psi}$ in the right panel. 
The gray shaded region 
($m_{\psi} \lesssim 0.45$ GeV for $\epsilon_2 = {0.005}$)
indicates the parameter space 
excluded by the millicharge constraints
\cite{Davidson:2000hf, Acciarri:2019jly, Ball:2020dnx}. 
}
\label{fig:compare-distribution}
\end{figure*}

In Fig.~\ref{fig:compare-distribution}, 
we compare  
the three dark photon production channels 
{(MD, PB, and HR)}
at the LHC, both in the $4\pi$ solid angle 
and in the very forward region.  
The very forward region is 
defined by the dark photon pseudorapidity $\eta_{A'} > 6$.\footnote{The 
angular acceptance of  FACET is $6< \eta <7.2$ \cite{FACET:talk1, Cerci:2021nlb}, 
and the angular acceptance of  FASER is $\eta > 9$ 
\cite{Ariga:2018zuc, Ariga:2018uku, Ariga:2018pin, Ariga:2019ufm}.}
We choose $\epsilon_{1} = 10^{-6}$ and $\epsilon_2 = {0.005}$ 
for both figures in Fig.~\ref{fig:compare-distribution}.

The left panel figure of Fig.~\ref{fig:compare-distribution} 
shows the dark photon cross sections as a function of the 
hidden fermion mass $m_\psi$ for the case 
where the dark photon mass is fixed at 
$m_{A'} \simeq 0.4$ GeV. 
The dark photon cross section in the HR process 
decreases with the hidden fermion mass $m_\psi$; 
the cross sections in the MD and PB processes 
are independent of $m_\psi$, 
since these processes do not involve the hidden fermion $\psi$. 
For light $\psi$ the HR process dominates the
MD and PB processes, 
whereas for heavy $\psi$ the MD and PB
processes become more important. 
In particular, the HR process dominates 
the dark photon production if 
$m_\psi \lesssim$ 5 GeV (30 GeV) in the 
very forward ($4\pi$ solid angle) region.

The right panel figure of Fig.~\ref{fig:compare-distribution} 
shows the dark photon cross sections as a function of the 
dark photon mass $m_{A'}$ for the case 
where $m_\psi = 5 \, m_{A'}$.  
The HR process dominates  
{the entire mass range except the small} 
resonance region near $m_{A'} \simeq 0.8$ GeV, 
where the PB process becomes larger. 
We note that, in the right panel of Fig.~\ref{fig:compare-distribution}, 
the resonance in the PB process is due to 
the pole structure (due to various vector mesons) 
in the form factor given
in Eq.~\eqref{eq:PB_formfactor}, 
and the kink features in the MD cross section 
arise because of the mass threshold
effects in meson decays.

About $10\%$ of the dark photons in the MD and PB processes
are produced in the very forward region as shown 
in Fig.~\ref{fig:compare-distribution}. 
For the HR process, 
the number of dark photons produced in the very forward region 
is sizable in the low $\psi$ mass region, 
with a fraction up to
$\sim {15\%}$ 
for $m_\psi \simeq 0.5$ GeV, 
as shown in the left panel figure of Fig.~\ref{fig:compare-distribution}. 
For heavy $\psi$ mass the cross section in 
the very forward region is significantly reduced, 
for example, less than 1\% of the dark photon 
in the HR process produced in the forward region
when $m_\psi \gtrsim 6$ GeV. 
This is because heavier $\psi$ particles tend to be produced more isotropically 
than lighter $\psi$ particles and thus lead to {fewer} 
events in the forward region.

\subsection{PDF uncertainties}

\begin{figure}[htbp]
\centering
\includegraphics[width=0.4\textwidth]{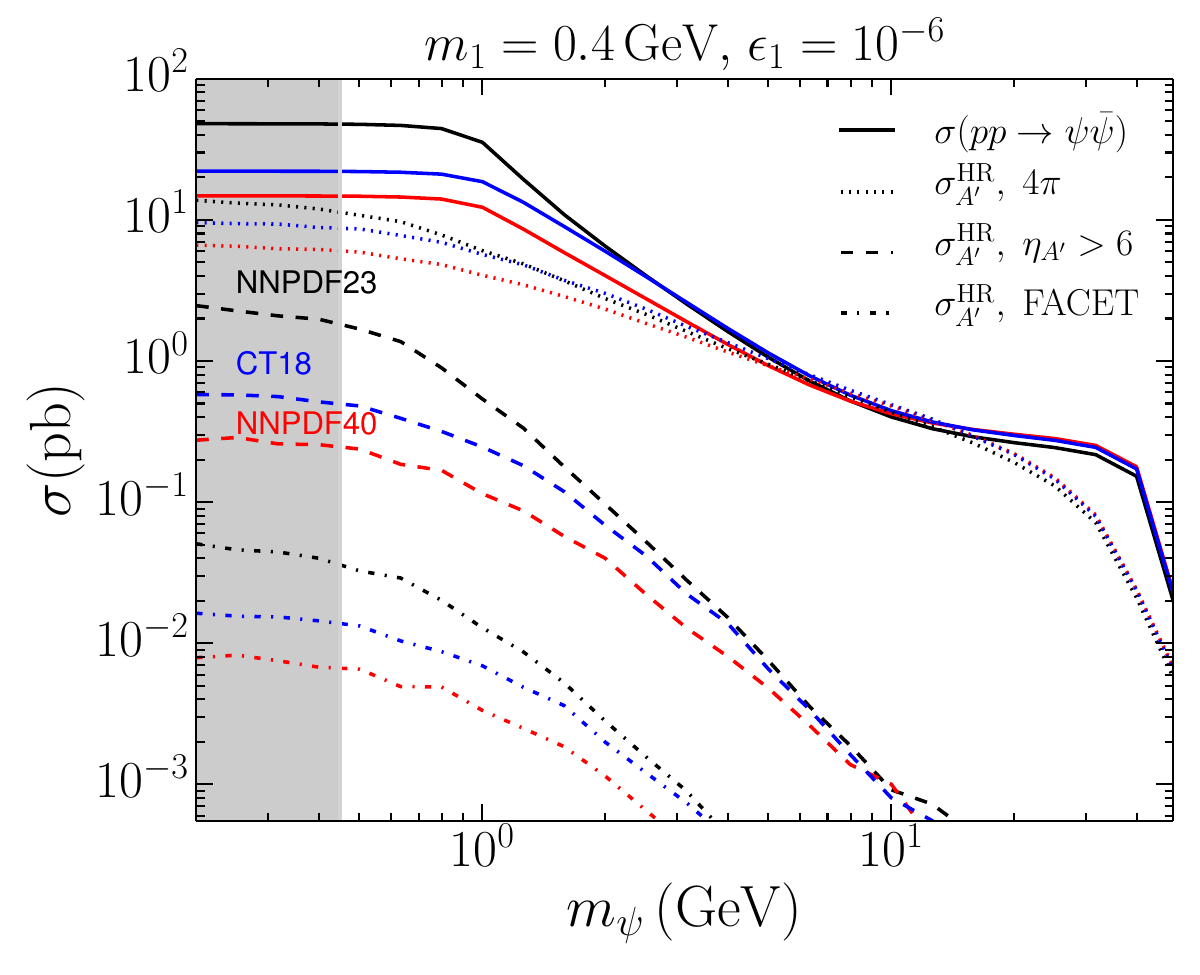}
\caption{Comparison of LHC cross sections using 
different PDFs. 
The LHC cross sections of   
$\sigma (pp\to \psi \bar\psi)$ (solid),
$\sigma(A')$ of the HR process 
in the $4 \pi$ angular region (dotted), 
in the forward region $\eta_{A'} > 6$ (dashed), 
and in the FACET detector ({dashdotted}) 
are computed with 
NNPDF23 \cite{Ball:2012cx} (black), 
NNPDF40 \cite{Ball:2021leu} (red), 
and CT18 \cite{Hou:2019qau} (blue).}
\label{fig:pdf-uncertainty}
\end{figure}

For light 
$\psi$ one has to integrate over the small $x$ region  
in PDFs where there are large uncertainties \cite{Feng:2017uoz}. 
In the process $pp \to \psi \bar \psi$, the minimum value of 
$x$ is  
$
x_{\rm min} = {4 m_\psi^2 / s}, 
$
if there is no cut on the $\psi$ momentum. 
Thus, for the $m_\psi = 15 \, (0.5)$ GeV case, 
one has to integrate over the $x$ range near  
$x_{\rm min} \simeq 5 \times 10^{-6}\, (6 \times 10^{-9})$. 
The minimum value of $x$ is $10^{-9}$ 
in the PDFs sets: 
NNPDF23LO \cite{Ball:2012cx}, 
NNPDF40 \cite{Ball:2021leu}, and 
CT18 \cite{Hou:2019qau}. 
Thus, for the $m_\psi = 0.5$ GeV case, 
the dark photon production cross section 
in the HR process (denoted as $\sigma_{A'}^{\rm HR}$) 
depends on 
the PDFs in the $x$ region where PDFs begin . 
To check the stability of the LHC cross sections 
(of small $m_\psi$)
against different PDFs, 
we compute various LHC cross sections 
including    
$\sigma (pp\to \psi \bar\psi)$,
$\sigma_{A'}^{\rm HR}$ 
in the $4 \pi$ angular region, 
$\sigma_{A'}^{\rm HR}$ 
in the forward region $\eta_{A'} > 6$, 
and 
$\sigma_{A'}^{\rm HR}$ 
in the FACET detector, 
by using three different PDFs: 
NNPDF23LO 
(the default PDFs in MADGRAPH 5), 
NNPDF40, and CT18, 
in Fig.\ \ref{fig:pdf-uncertainty}.

For $\sigma(pp \to \psi \bar\psi)$ 
at $m_\psi \simeq 0.5$ GeV,  
the NNPDF40 (CT18)
leads to a cross section that is about 
{$30\%$ ($45\%$)} 
of that from NNPDF23;  
for $\sigma_{A'}^{\rm HR}$ 
in the $4 \pi$ angular region, 
these two percentage numbers become 
{$55\%$ ($80\%$)}.
This is because the $\psi$ particles have to be 
energetic enough to radiate 
dark photons, 
{which then corresponds} to 
larger $x_{\rm min}$ values in the PDF integration,  
leading to less PDF uncertainties. 
The  PDF uncertainties in the $4\pi$ angular region 
are smaller than the forward region, which is 
due to the fact that the $4\pi$ region includes 
the region with significant transverse momentum.

In the sensitivity contours of FACET as shown in 
Fig.\ \ref{fig:psi5DP}, 
the mass of $\psi$ has to satisfy $m_\psi \gtrsim 1.5$ GeV 
to be consistent with the millicharge constraints. 
We find that NNPDF40 (CT18) leads to a cross section 
of $\sim 33\%$ ($\sim 64\%$) of NNPDF23 
at $m_\psi \simeq 1.5$ GeV, as shown 
in Fig.\ \ref{fig:pdf-uncertainty}. 
For the $m_\psi \simeq 15$ GeV case 
(the $\psi$ mass in Fig.\ \ref{fig:psi15GeV}), we find that 
the cross section computed with NNPDF40 (CT18) 
is $\sim 80\%$ ($\sim 97\%$) of that with NNPDF23, 
in Fig.\ \ref{fig:pdf-uncertainty}.
Thus the PDF uncertainty on our sensitivity contours  
is less significant. Furthermore, 
the sensitivity contours analyzed with different 
PDFs, as shown in  Fig.\ \ref{fig:psi5DP}, 
show that different PDFs only modify the limits  
for small $\epsilon_1$ values (the lower edge of 
the contours), but have unnoticeable effects on 
{large} $\epsilon_1$ values (the upper edge of 
the contours). 
This is due to the fact that 
the large $\epsilon_1$ values correspond to small 
decay lengths, and thus the dark photon should have a 
significant momentum to decay inside the far detectors. 
For that reason, the $x_{\rm min}$ in the PDF integration 
becomes larger for 
the model points with large $\epsilon_1$ 
values, resulting in insignificant PDF uncertainty.

\section{Analysis}
\label{sec:simu-and-considerations}

In this analysis, we investigate the LLDP 
signals in the following four detectors: 
FACET, FASER, MATHUSLA, 
and CMS-MTD. 
We carry out analysis for the model points 
in the parameter space spanned by  
the DP mass $m_{A'}$ and the DP lifetime $\tau_{A'}$.
For each model point, 
we compute the DP signal events  
from the MD, PB and HR processes.
For the MD and PB processes,
we obtain the DP momentum 
and the position of its decay vertex, 
by using the simulations discussed in 
section \ref{subsec:MD} and 
section \ref{subsection:proton_bremss}, respectively.
We then boost the daughter particles from 
dark photon decays to the lab frame, 
from the rest frame of the dark photon, 
where the daughter particles are isotropic.
For the HR process, we use MADGRAPH 5 \cite{Alwall:2014hca} 
to generate $10^6$ events for the 
$p p \to \psi \bar{\psi}$ process, and use PYTHIA 8 
\cite{Sjostrand:2014zea, Carloni:2010tw, Carloni:2011kk} 
to simulate the hidden radiation of the $\psi$ particle 
and the decay of the dark photon, 
which outputs the momentum information 
for the DP and its daughter 
particles, as well as the decay position 
of the DP. 
To expedite the analysis  
(only a small fraction of simulated events from PYTHIA 8 
are actually inside the decay volume of the detectors), 
we disregard the decay position 
of the dark photon provided by 
PYTHIA 8 and use the dark photons 
that decay both inside and outside of the decay volume.

Thus, for the three far detectors (FACET, FASER, MATHUSLA), 
we compute the probability of detecting a DP as follows 
\be
P_{A'} = 
f(\theta, \phi) 
\int_{L_{\rm min}}^{L_{\rm max}} d \ell 
\frac{e^{-\ell/\ell_{A'}}} {\ell_{A'}} \, \omega \, , 
\label{eq:prob-detection}
\ee
where $L_{\rm min}$ ($L_{\rm max}$)
is the minimal (maximum)
distance between the decay volume and the IP 
along the $(\theta, \phi)$ direction 
{with $\theta$ and $\phi$ the polar and azimuthal angles of 
the dark photon respectively}, 
$\ell_{A'} = \tau_{A'} |\vec{p}_{A'}|/m_{A'}$ 
is the decay length of dark photon 
with $\tau_{A'}$ being the lifetime, 
$f(\theta, \phi)$ describes the angular acceptance 
of the decay volume,
and 
{$\omega$ equals 1 if the decay final states of the
DP satisfy
additional detector cuts
($\omega$ equals 0 otherwise).}

For a cylindrical detector (e.g.\ FASER and FACET)
that is 
placed along the beam direction 
with a distance $d$ from the IP 
to the near side of the detector, 
the parameters in Eq.~\eqref{eq:prob-detection}  are given by
\bea
L_{\rm min} &=& d, \quad 
L_{\rm max} = d+L, \\
f(\theta, \phi) &=& \Theta(R/{L_{\rm min}} - \tan \theta) \,\Theta(\tan \theta - r/L_{\rm max}),
\label{eq:fthetaphi}
\eea
where 
$L$ is the length of decay volume of the detector, 
$r$ ($R$) is the inner (outer) radius of the decay volume, 
and $\Theta$ is the Heaviside step function. 
For the FACET detector, one has  
$r={18}$ cm and $R=50$ cm; 
for the FASER (FASER 2) detector,  one has $r=0$ and $R=10$ (100) cm. 
For the cylindrical forward detectors, 
the pseudorapidity range is often used 
to describe the acceptance of the detectors, 
$f(\theta, \phi) = \Theta(\eta_{\rm{max}} - \eta_{A'}) \Theta(\eta_{A'} - \eta_{\rm{min}})$. 
Thus for the FACET detector, 
one has $\eta_{\rm{min}} \simeq 6$ and 
$\eta_{\rm{max}}\simeq {7.2}$;\footnote{$\eta_{\rm{min}} \simeq 6$ 
corresponds to 
the left-upper corner of the decay volume, and 
$\eta_{\rm{max}}\simeq {7.2}$ corresponds to 
the right-bottom (inner radius) corner 
of the upper half of the decay volume as shown in Fig.~\ref{fig:facet-layout1}.
}
for the FASER (FASER 2) detector, 
one has $\eta_{\rm{min}} \simeq 9$ (7) and $\eta_{\rm{max}} = +\infty$.

For a box-shape detector with height $H$, width $W$, length $L$
and is located at a distance $d$ from the IP along the z-axis
and a distance $h$ above the LHC beam (along the x-axis) 
(e.g.\ MATHUSLA), 
one has\footnote{Note the distance $d$ here is different from 
that in Table \ref{tab:detectors}.} 
\begin{widetext}
\bea
\label{eq:box1-new}
 L_{\rm max} &=& \left\{
\begin{aligned}
&\frac{h+H}{\sin \theta \cos \phi}\quad& {\rm if} \;\;
& \tan \theta > \frac{h + H } { (d+L) \cos \phi}\; \&\; |\tan\phi | < \frac{W}{2(h+H)},\\
&\frac{d+L}{\cos \theta }\quad&  {\rm if} \;\; 
&  \tan \theta < \frac{h + H } { (d+L) \cos \phi}\; \&\; |\sin\phi | < \frac{W}{2(d+L)\tan\theta}, \\
&\frac{W}{2\sin \theta |\sin\phi|}\quad&  {\rm if} \;\; 
& |\sin\phi | > \frac{W}{2(d+L)\tan\theta},
\end{aligned}
\right.  \\
 \label{eq:box2-new}
L_{\rm min} &=& \left\{
\begin{aligned}
&\frac{ h }{\sin \theta \cos \phi}\quad& {\rm if} \;\; & \tan \theta < \frac{h } { d \cos \phi}, \\
&\frac{d}{\cos \theta }\quad&  {\rm if} \;\; &  \tan \theta > \frac{h} {d\cos \phi},
\end{aligned}
\right.  \\
\label{eq:box3-new}
f(\theta, \phi) &=&  \Theta \left( \tan \theta - \frac{h}{(d+L)\cos \phi} \right) \, 
 \Theta \left(  \frac{h + H }{d\cos \phi} -  \tan \theta\right) \, 
 \Theta \left(\frac{W}{2 h} - |\tan \phi| \right)  \Theta \left(\cos\phi \right). 
\eea
\end{widetext}
For the MATHUSLA detector, we use 
$d=$ 68 m, 
$h=$ 60 m, 
$W=$ 100 m, 
$L=$ 100 m, 
and $H=$ 25 m 
\cite{Alpigiani:2020tva}.

For FACET,
we {further} require both daughter particles from the DP decay to
traverse both the tracker and 
the calorimeter detectors. 
 For {the} FASER detector, we further apply a detector cut
 on the energy of DP daughter particles 
 $E_{\rm vis} > 100$ GeV \cite{Ariga:2018uku}
to reduce the trigger rate 
and remove possible {background} (BG) at low energies.
For {the} FACET detector, because the BG events
are expected to be highly suppressed  
due to the front shielding and
the high quality vacuum of the decay volume, 
no detector cut is required. 
For {the} MATHUSLA  detector, we require both
DP daughter particles to hit the ceiling detector 
and are well separated with an opening angle 
$\Delta \theta > 0.01$ \cite{Curtin:2018mvb}; 
we note that $\omega=0$ 
for the second and third lines of
Eq.~\eqref{eq:box1-new}, by requiring such a cut.

Thus the number of events in the far detector can be obtained 
\be
N = {\cal L} \cdot \sigma_{A'} \cdot \langle P_{A'} \rangle  \quad {\rm with }  \quad
\langle P_{A'} \rangle = \frac{1}{N_{\rm A'}} \sum_{i=1}^{N_{A'}} P_{A'_i}, 
\ee
where $\sigma_{A'}$ is the total DP production cross section,
$\langle P_{A'} \rangle$ denotes the average detection probability 
of the DP event, 
$N_{A'}$ is the total number of the DP in the simulation
and $P_{A'_i}$ is the individual detection probability of 
the $i$th dark photon event in the simulation 
which is given by Eq.~\eqref{eq:prob-detection}.

For the CMS-MTD detector, we only
consider the DPs produced from the HR 
process for the CMS-MTD analysis. This is because 
the CMS-MTD detector does not have sensitivity  
to the DP mass below $\sim$GeV \cite{Du:2019mlc}. 
Following {Ref.\ \cite{Du:2019mlc}},
we use MADGRAPH 5 
to generate $\psi \bar \psi$ events 
with an ISR jet to time stamp the event, i.e., 
$pp \to \psi\bar{\psi}j$ where 
the ISR jet is required to have $p_T > 30$ GeV and $|\eta| < 2.5$.
The DP is required to have 
a transverse decay length $0.2 {\rm m} < \ell_{A'}^T < 1.17 {\rm m}$ 
and a longitudinal decay length $|z_{A'}| <3.04$ m. 
The final state leptons from DP decays 
are detected by the precision timing detector; 
the leading lepton should have $p_T > 3$ GeV. 
The time delay variable \cite{Liu:2018wte}
between the ISR jet and the leading lepton 
is required to 
$\Delta t > 1.2\ \rm{ns}$ \cite{Du:2019mlc}.

\section{Result}
\label{sec:result}

\begin{figure*}[htbp!]
\begin{centering}
\includegraphics[width=0.4\textwidth]{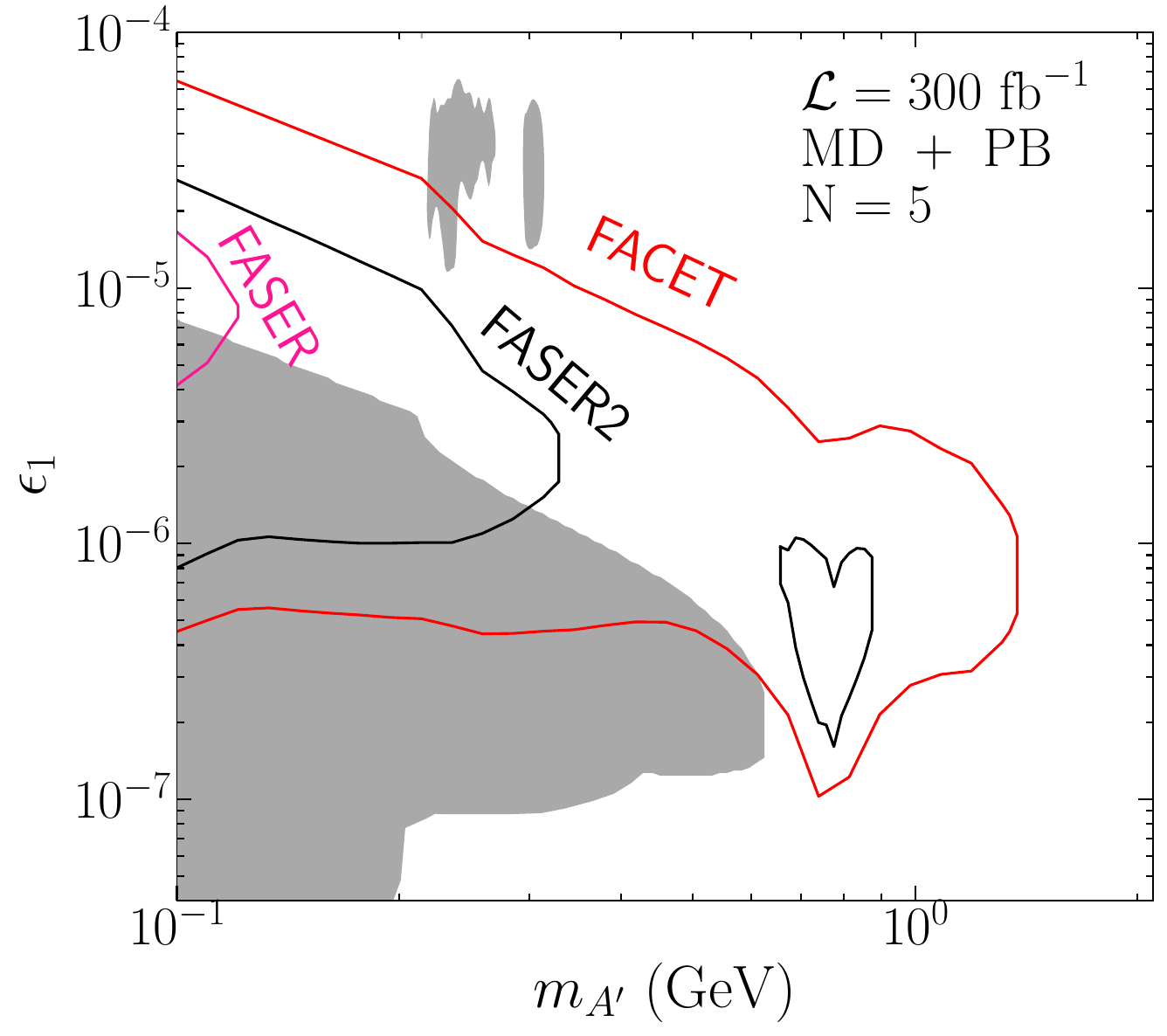}
\includegraphics[width=0.4\textwidth]{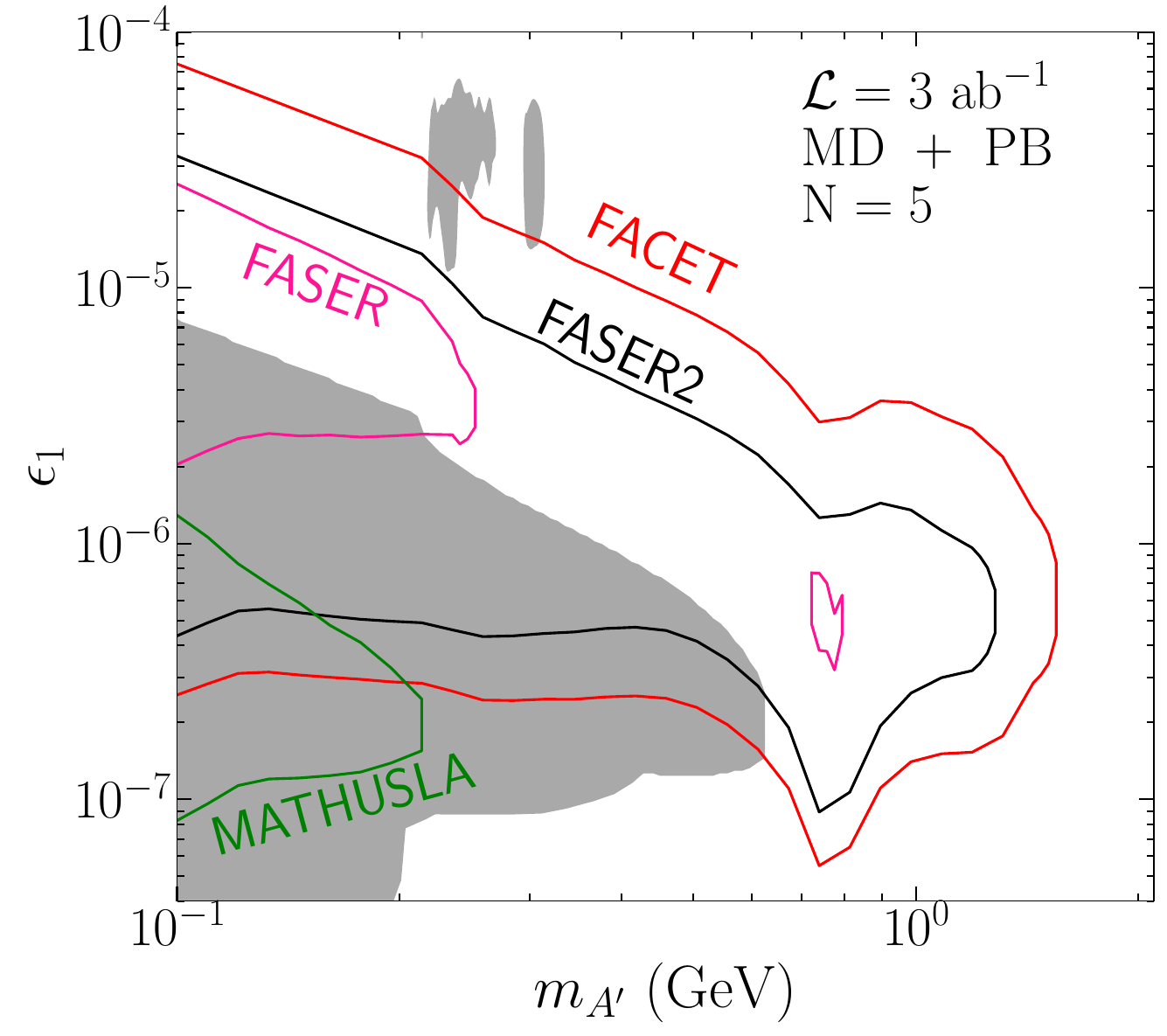}
\caption{Projected sensitivities from 
FACET (red), 
FASER (magenta), 
FASER2 (black), 
and MATHUSLA (green), 
at the HL-LHC 
with the integrated luminosities of 
${\cal L} = 300$ fb$^{-1}$ (left panel) 
and ${\cal L} = 3$ ab$^{-1}$ (right panel)
to the ``minimal'' dark photon models 
in which only the MD and PB processes 
contribute to the signals. 
Contours correspond to 
the expected signal events $N=5$.   
The dark gray shaded region indicates the 
parameter space that has been excluded 
by various experiments 
{including LHCb~\cite{LHCb:2019vmc}, 
$\nu$-CAL~I~\cite{Blumlein:2011mv,Blumlein:2013cua},
CHARM~\cite{Gninenko:2012eq} 
and E137~\cite{Bjorken:1988as}}; 
the limits are obtained with the Darkcast 
package \cite{darkcast}. 
}
\label{fig:MDplusPB} 
\end{centering}
\end{figure*}

\begin{figure*}[htbp]
\begin{centering}
\includegraphics[width=0.4\textwidth]{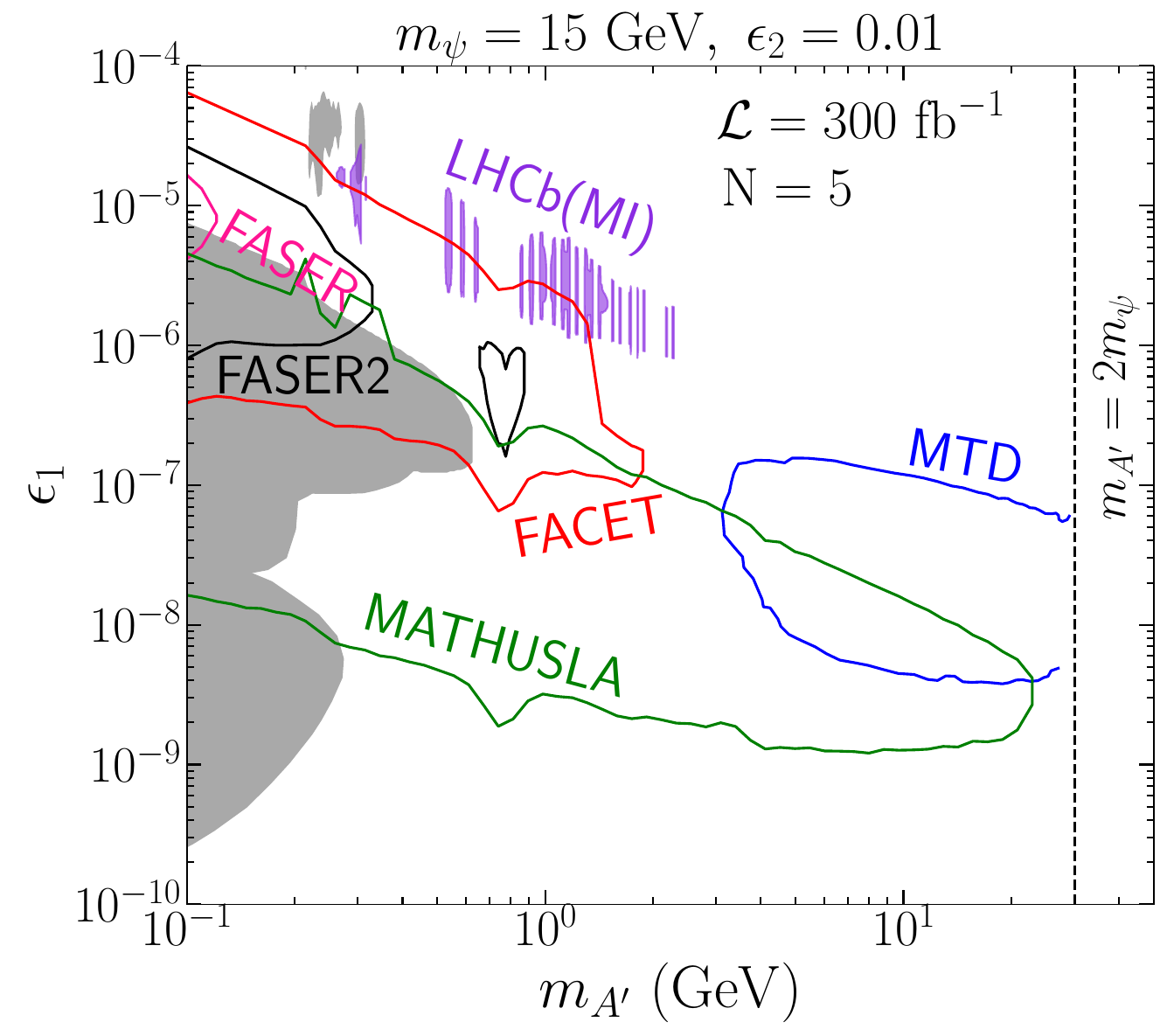}
\includegraphics[width=0.4\textwidth]{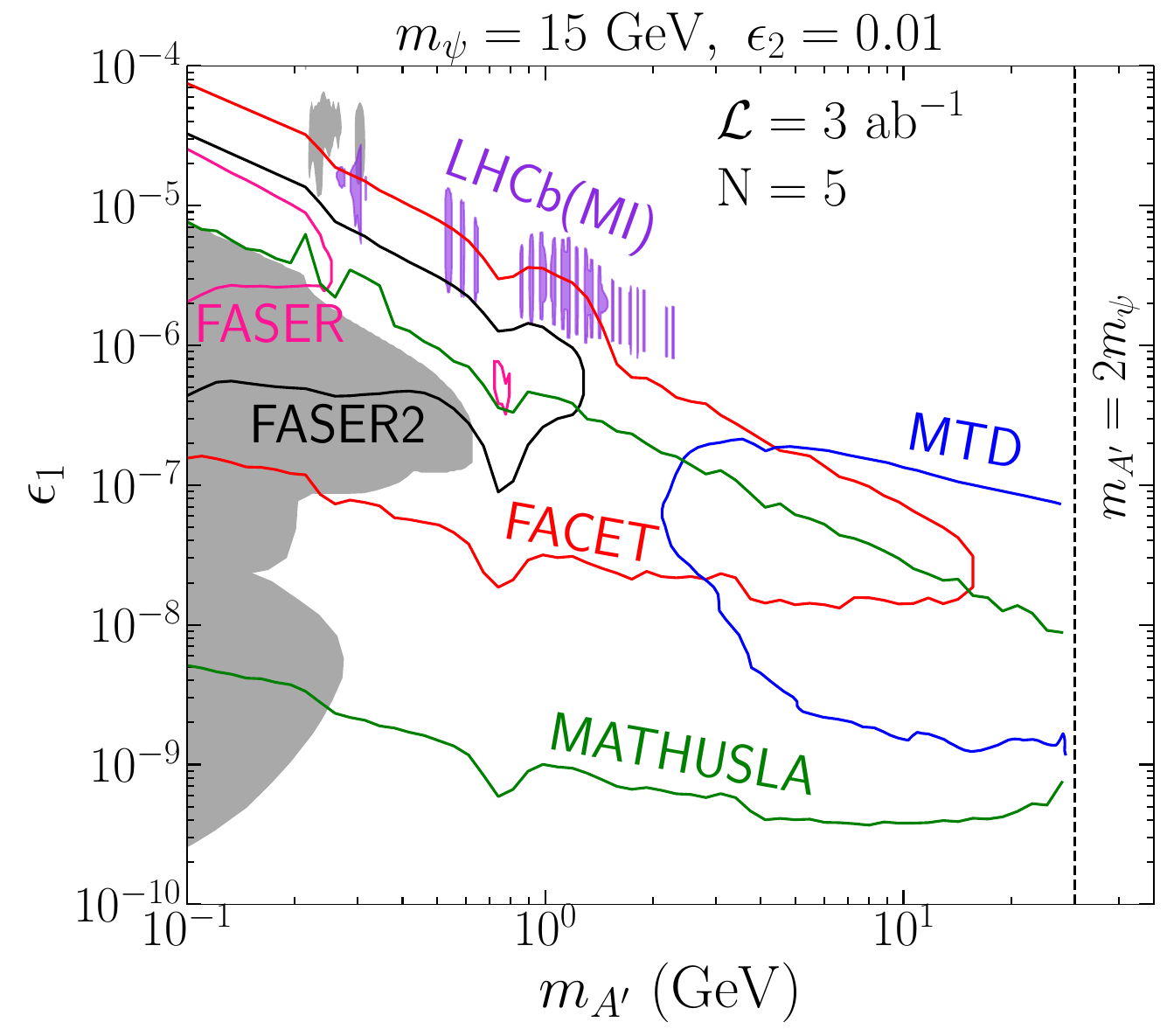}
\caption{Projected sensitivities from 
FACET (red), 
FASER (magenta), 
FASER2 (black), 
and MATHUSLA {(green)}, 
at the HL-LHC 
with the integrated luminosities of 
${\cal L} = 300$ fb$^{-1}$ (left panel) 
and ${\cal L} = 3$ ab$^{-1}$ (right panel)
to our dark photon model  
in which all the three dark photon 
production channels 
(MD, PB, and HR) 
contribute to the signals. 
Here we fix $m_\psi = 15$ GeV 
and $\epsilon_2 = 0.01$, and 
require $m_{A'} < 2 m_{\psi}$ 
so that the dark photon cannot 
decay into invisible final states.
Contours correspond to 
the expected signal events $N=5$.   
The dark gray shaded region indicates the 
excluded dark photon parameter space 
by various experiments {including LHCb~\cite{LHCb:2019vmc}, 
$\nu$-CAL~I~\cite{Blumlein:2011mv,Blumlein:2013cua},
CHARM~\cite{Gninenko:2012eq}, 
E137~\cite{Bjorken:1988as}},
{LSND~\cite{Athanassopoulos:1997er},
and SN1987A~\cite{Chang:2016ntp}}
where the HR process is not considered; 
the limits are obtained with the Darkcast 
package \cite{darkcast}. 
The purple shaded {regions}  
are excluded 
by recasting 
the model-independent (MI) constraints  
from the displaced di-muon search 
at the LHCb \cite{LHCb:2020ysn} 
on the HR process.}
\label{fig:psi15GeV} 
\end{centering}
\end{figure*}

\begin{figure*}[htbp]
\begin{centering}
\includegraphics[width=0.4\textwidth]{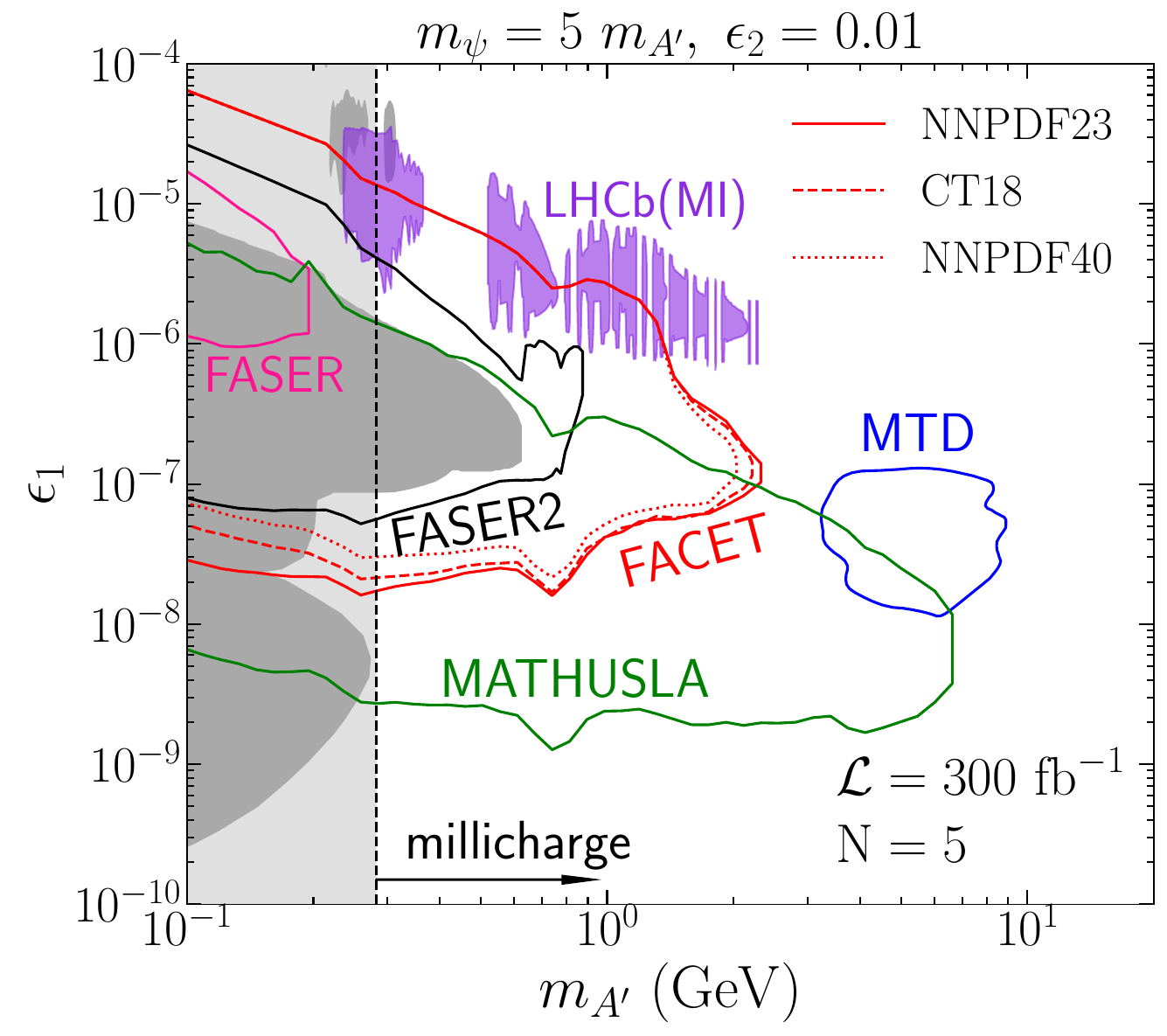}
\includegraphics[width=0.4\textwidth]{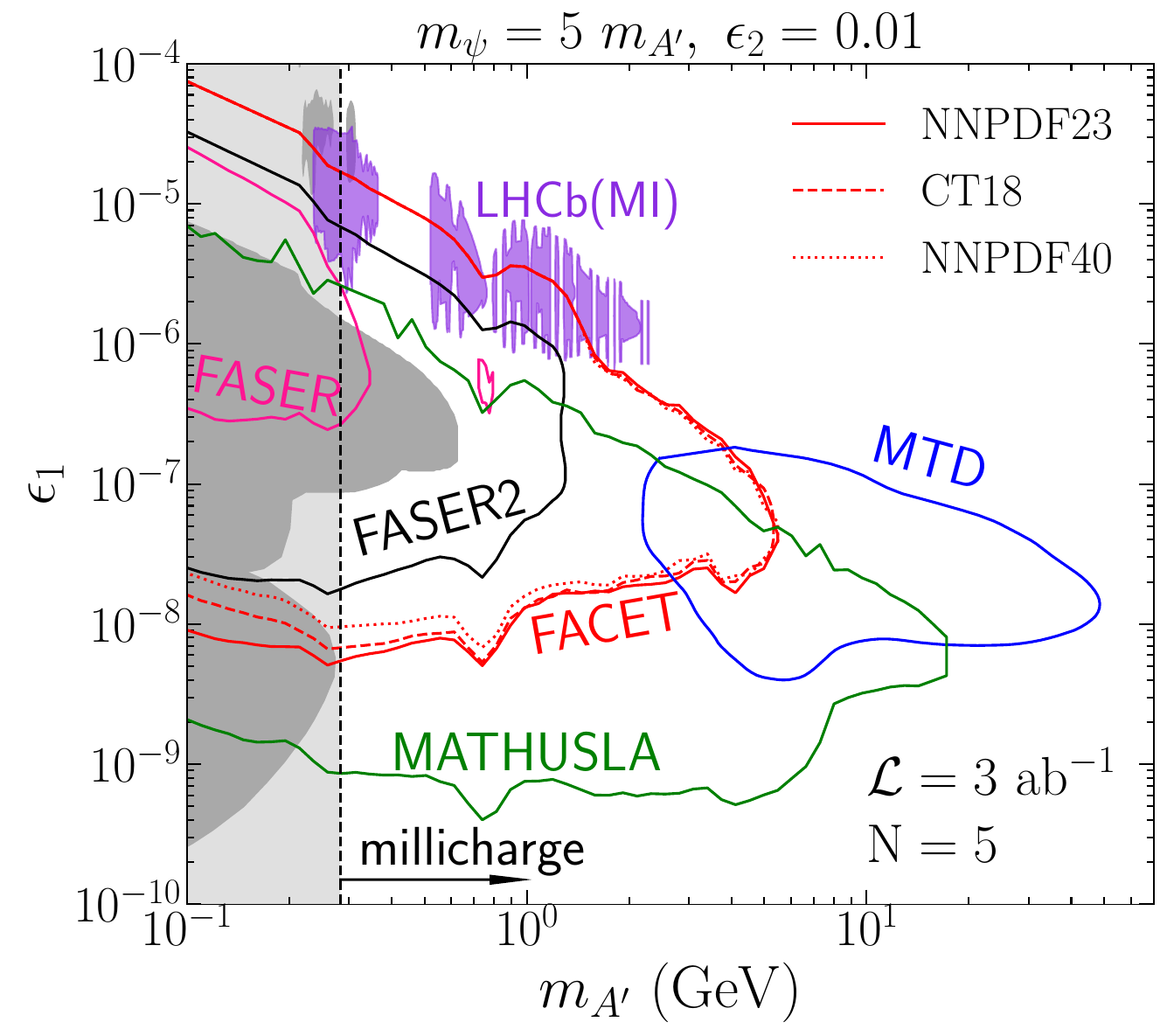}
\caption{Same as Fig.\ \ref{fig:psi15GeV} except 
$m_\psi=5 \, m_{A'}$.  
The light gray region is excluded by 
the millicharge constraints 
\cite{Davidson:2000hf, Acciarri:2019jly, Ball:2020dnx}. 
For the FACET contours we use  
NNPDF23 \cite{Ball:2012cx} (red-solid), 
CT18 \cite{Hou:2019qau} (red-dashed),
and NNPDF40 \cite{Ball:2021leu} (red-dotted).}
\label{fig:psi5DP} 
\end{centering}
\end{figure*}

\begin{figure*}[htbp]
\begin{centering}
\includegraphics[width=0.41\textwidth]{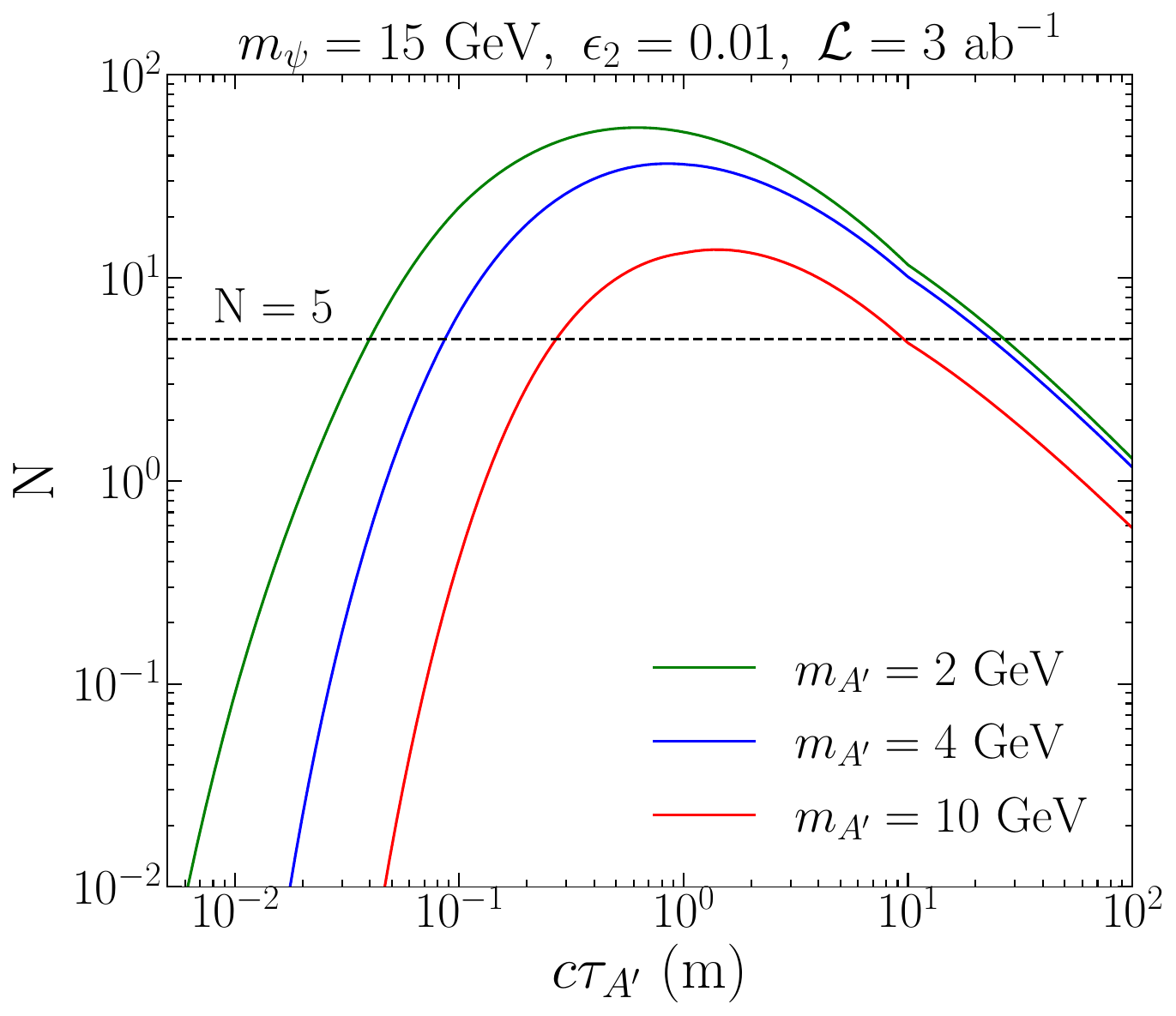}
\includegraphics[width=0.4\textwidth]{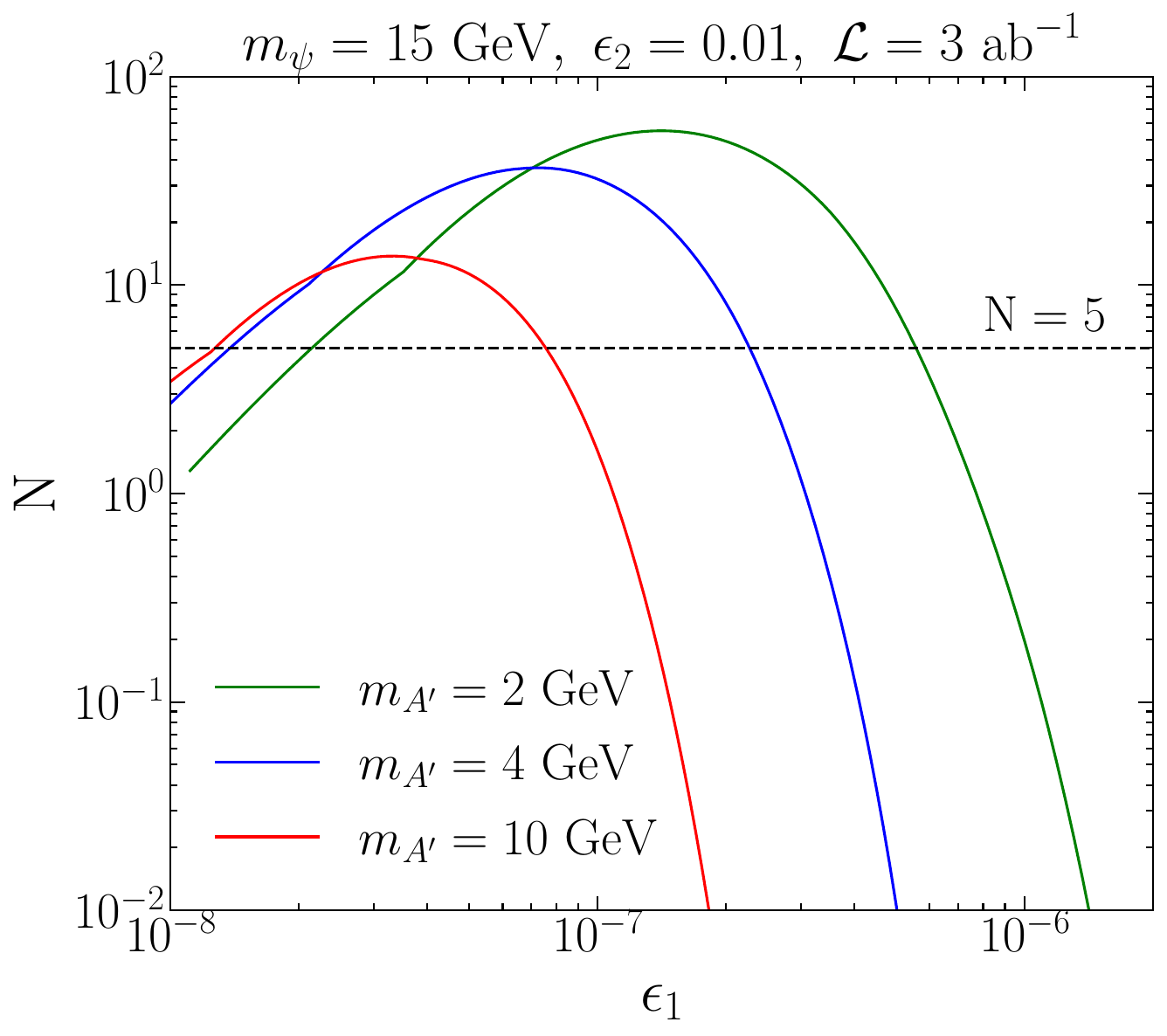}
\caption{The number of signal events 
in the FACET detector
at the HL-LHC with ${\cal L} = 3 \,{\rm ab}^{-1}$, 
as a function of
the DP lifetime {$c\tau_{A'}$} (left panel)
and of the coupling $\epsilon_1$ (right panel). 
Here we fix $m_\psi = 15$ GeV and $\epsilon_2 = 0.01$, 
and vary the dark photon mass to be 
$m_{A'} = 2$ GeV (green), 
4 GeV  (blue), 
and 10 GeV (red).
}
\label{fig:FACET-Nevent} 
\end{centering}
\end{figure*}

In this section we discuss the projected sensitivities of 
the future LLP detectors including FACET, FASER, MATHUSLA 
and the precision timing detector {CMS-MTD}. 
Our main results are shown in 
Figs.\ (\ref{fig:MDplusPB}, 
\ref{fig:psi15GeV}, 
\ref{fig:psi5DP}, 
\ref{fig:FACET-Nevent}), 
{where sensitivity contours for far detectors 
are made by requiring the new physics events 
to be $N = 5$, under the assumption that 
the SM processes do not contribute any event 
in the decay volume after various shieldings 
and detector cuts.} 
We are only interested in the parameter space in which $m_{A'} < 2 m_\psi$ 
so that the dark photon {is} kinematically forbidden 
to decay into the hidden fermion pair, {leading to a 
long-lived dark photon.}\footnote{{If 
$m_{A'} > 2 m_\psi$, the dark photon can decay into 
a pair of hidden fermions, which then leads to a 
prompt decay dark photon, assuming an order-one 
gauge coupling in the hidden sector.}}

Fig.\ \ref{fig:MDplusPB} shows the projected sensitivities 
on the minimal dark photon models 
with $300$ fb$^{-1}$ 
and $3$ ab$^{-1}$ data, 
from 
FACET, FASER, FASER2, and MATHUSLA. 
We also exhibit various experimental constraints   
including LHCb~\cite{LHCb:2019vmc}, 
$\nu$-CAL~I~\cite{Blumlein:2011mv,Blumlein:2013cua},
CHARM~\cite{Gninenko:2012eq} 
and E137~\cite{Bjorken:1988as}.
We only include the MD and PB processes here; 
the HR process is absent. 
For that reason, the analysis in Fig.\ \ref{fig:MDplusPB} 
is also applicable to the minimal dark photon model.
Among the new detectors, 
the parameter space probed by 
FACET is larger than the
other experiments. 
In particular, with an integrated luminosity of 
$ 300\, {\rm fb}^{-1}$ (3 ${\rm ab}^{-1}$) at the HL-LHC,
FACET can probe the DP mass up to $\sim 1.3$ GeV 
{($1.5$ GeV)}, 
whereas FASER can only probe 
the DP mass up to $\sim 0.12$ GeV 
{($0.25$ GeV plus the island near $0.79$ GeV)},
and FASER2 can only probe 
the DP mass up to $\sim 0.8$ GeV 
{($1.3$ GeV)}.
Because DPs arising from the PB and MD processes are likely
to be distributed in the forward region,
MATHUSLA, a detector located in the central transverse region,
has difficulties to probe the parameter space of the minimal dark photon model. 
For that reason, MATHUSLA only probe a small parameter region 
with 3 ${\rm ab}^{-1}$ data, 
which, however, has been excluded already 
by the current experimental constraints. 
We note that the dips at $m_{A'} \sim 0.8$ GeV in the contours 
are due to the resonance in the PB process, 
and the kink features 
at $m_{A'} \sim 0.2$ GeV are due to
the mass threshold effects in the MD process.

Fig.~\ref{fig:psi15GeV} shows the projected sensitivities 
for our dark photon models 
{from FACET, FASER, FASER2, MATHUSLA, 
and CMS-MTD.} 
Here the dark photon production contributions 
from all channels including the MD, PB and HR processes 
are considered.   
With the inclusion of the HR process, 
the FACET and MATHUSLA sensitivity 
contours are significantly enlarged 
to heavier DP mass region, 
as compared to Fig.~\ref{fig:MDplusPB}; 
the FASER and FASER2 sensitivity contours, 
on the other hand, are similar to those in Fig.~\ref{fig:MDplusPB}.
With $ 300\, {\rm fb}^{-1}$ ($3\,{\rm ab}^{-1}$) data at the HL-LHC, 
FACET can probe the parameter space of our model 
up to $ m_{A'} \simeq 1.9 \, (15)$ GeV. 
The {CMS-MTD} probes a relative large dark photon mass region: 
down to dark photon mass  $\sim 3 \, (2)$ GeV 
for $ 300\, {\rm fb}^{-1}$ ($ 3\, {\rm ab}^{-1}$) data at HL-LHC. 
This is due to the fact that a light dark photon
leads to not only a small time delay
but also small transverse momenta
of the final state leptons,
which will suffer from a large SM background
for the time delay searches \cite{Du:2019mlc}.
Interestingly, this {CMS-MTD} sensitivity region partly overlaps 
with MATHUSLA sensitivity region for the luminosity of $ 300\, {\rm fb}^{-1}$, 
and with both FACET and MATHUSLA sensitivity regions 
for the luminosity of $ 3\, {\rm ab}^{-1}$. 
Thus, if a dark photon in this overlap region is discovered, 
one can {see} 
the FACET and MATHUSLA to 
verify the results from the CMS-MTD.

{Fig.~\ref{fig:psi5DP} 
shows the expected limits  
from FACET, FASER, FASER2, MATHUSLA, 
and CMS-MTD 
to the parameter space of our dark photon model 
with the mass relation $m_\psi=5 \, m_{A'}$.}  
{The sensitivity contours are similar to Fig.~\ref{fig:psi15GeV}, 
but with some changes.} 
{For light $\psi$, the millicharge constraints are important, 
which excludes the parameter space $m_{A'} \lesssim 0.3$ GeV 
(corresponding to $m_\psi > 1.5$ GeV for $\epsilon_2 = 0.01$).} 
The parameter space probed by FASER with 
${\cal L} = 300\, {\rm fb}^{-1}$ ($3\, {\rm ab}^{-1}$) at the HL-LHC 
is (nearly) excluded by the millicharge constraints. 
Further, the heavy dark photon mass region can no longer be 
probed by various detectors as in Fig.~\ref{fig:psi15GeV}. 
This is because the heavy dark photon mass corresponds to the 
heavy $\psi$ mass via the mass relation $m_\psi=5 \, m_{A'}$, 
which then leads to a suppressed $pp \to Z^* \to \psi \psi$ cross section. 
Similar to the result in Fig.~\ref{fig:psi15GeV},
the {CMS-MTD} sensitivity region is  
partly overlapped with FACET and MATHUSLA. 
To check the PDF uncertainties on the sensitivity contours, 
we further compute the FACET contours 
using three different sets of PDFs: 
NNPDF23 \cite{Ball:2012cx} (red-solid), 
CT18 \cite{Hou:2019qau} (red-dashed),
and NNPDF40 \cite{Ball:2021leu} (red-dotted). 
As shown in Fig.~\ref{fig:psi5DP}, 
the upper edge of the FACET contours 
from the three PDFs are almost identical; 
the lower edge of the FACET contours 
from the three PDFs, however, 
 can be seen with some visible differences 
 from each other. 
For example, 
for $m_{A'} \sim 0.3$ GeV,
the lower edge of the FACET contours with 
$ 300\, {\rm fb}^{-1}$
are located at 
$\epsilon_1 = 1.9 \times 10^{-8}$ with NNPDF23, 
$\epsilon_1 = 2.3 \times 10^{-8}$ with CT18, 
and $\epsilon_1 = 3.2 \times 10^{-8}$ with NNPDF40, 
as shown on the left panel figure of Fig.~\ref{fig:psi5DP};  
for $3\, {\rm ab}^{-1}$ data, $\epsilon_1$ are 
$5.9 \times 10^{-9}$,
$7.3 \times 10^{-9}$, 
and $1.0 \times 10^{-8}$  respectively, 
as shown on the right panel figure of Fig.~\ref{fig:psi5DP}. 
Thus different PDFs will result in changes 
to the FACET contours but the effects are not significant.

Model-independent constraints on LLPs with 
a displaced vertex of several centimeters 
in the di-muon channel have been recently 
analyzed by LHCb  \cite{LHCb:2020ysn}, 
which used the same data sample ($5.1\,\rm{fb}^{-1}$) as the 
analysis optimized for the 
minimal dark photon model \cite{LHCb:2019vmc}, 
but with a different fiducial region and selection cuts. 
The 90\% CL upper bounds on 
the LLP cross section $\sigma(X\to \mu^+\mu^-)$ 
are provided  
for various LLP masses in the 
range $0.214$  GeV $<m_X<3$ GeV and for three 
$p_T^X$ bins: 2-3 GeV, 3-5 GeV, 
and 5-10 GeV \cite{LHCb:2020ysn}. 
This allows us to recast the limits to the 
HR process in our model.\footnote{The MD and 
PB processes in our model is the same as the 
minimal DP model. According to Ref.\ 
\cite{LHCb:2020ysn}, 
the new LHCb analysis 
\cite{LHCb:2020ysn} 
is only half sensitive in probing the minimal DP model, 
as compared with the previous LHCb analysis  
\cite{LHCb:2019vmc}. 
Thus, we exclude the MD and PB 
processes in our simulation and only consider the HR process.}
Following Ref.\ \cite{LHCb:2020ysn}, 
we select events with muon transverse momentum
$p_T(\mu) > 0.5$ GeV, muon momentum
$10 < p(\mu) < 1000$ GeV, 
muon pseudorapidity 
$2 < \eta(\mu) < 4.5$, 
and 
${p_T(\mu^+) \, p_T(\mu^-)} > 1$ GeV$^2$. 
We further require that 
the DP has a transverse momentum of $2 < p_T(A') < 10$ GeV, 
a pseudorapidity of $2 < \eta (A') < 4.5$, 
and a transverse decay length of $12 < \ell_{A'}^T < 30$ mm, 
and the opening angle of the di-muon pair is larger than 3 mrad. 
We then rule out a model point in the parameter space 
if it produces a cross section exceeding the upper bound 
in any of the three $p_T$ bins. 
We show the excluded parameter space by this 
model-independent LHCb 
analysis (purple shaded regions) in 
Figs.~\ref{fig:psi15GeV} and~\ref{fig:psi5DP}. 
For the minimal dark photon, the excluded regions are 
the two {gray} ``islands'' at $\sim$0.2-0.3 GeV \cite{LHCb:2019vmc}. 
For our dark photon model, the excluded regions  
by the LHCb model-independent limits 
are much larger,  
extending 
beyond $\sim$2 GeV in the dark photon mass 
and 
down to {$\epsilon_1 \sim 10^{-6}$}, 
which already rule out some parameter space 
to be probed by FACET and FASER2 detectors.

The left panel in Fig.~\ref{fig:FACET-Nevent} shows
the number of signal events in the FACET detector 
as a function of the proper lifetime, for three 
different dark photon masses.   
The number of events {decreases} 
with the dark photon mass. 
The peak of the distribution of the events shifts {to} a 
larger $c\tau_{A'}$ value when the dark photon mass increases. 
The peak shift is due to the detector cut on the DP decay length: 
a larger $c \tau_{A'}$ is needed for a heavier DP mass  
so that the DP has the desired decay length 
to disintegrate in the FACET decay volume.
With the criterion of $N > 5$ events, FACET 
can probe the $c\tau_{A'}$ range of $[0.04\, {\rm m} - 30 \, {\rm m}]$ 
for DP mass $m_{A'} = 2$ GeV, 
$c\tau_{A'} \in [0.09\, {\rm m} - 25 \, {\rm m}]$ 
for $m_{A'} = 4$ GeV, 
and $c\tau_{A'} \in [0.3\, {\rm m} - 10 \, {\rm m}]$ 
for $m_{A'} = 10$ GeV.
The right panel in Fig.~\ref{fig:FACET-Nevent} 
shows the number of signal events 
in the FACET detector as a function 
of the parameter $\epsilon_1$. 
With the criterion of $N > 5$ events, FACET 
can probe the 
{$\epsilon_1 \in [2.1 \times 10^{-8} - 5.5 \times 10^{-7}]$}
for DP mass $m_{A'} = 2$ GeV, 
{$\epsilon_1 \in [1.4 \times 10^{-8} - 2.3 \times 10^{-7}]$}
for $m_{A'} = 4$ GeV, 
and 
{$\epsilon_1 \in [1.3 \times 10^{-8} - 7.5 \times 10^{-8}]$}
for $m_{A'} = 10$ GeV.

\section{Expected number of events in far detectors}
\label{sec:facet-faser-comparison}

Here we provide an approximated expression for the number 
of dark photon events in the far detectors, and also compare 
the number of events for two far detectors 
that are of different sizes and placed with different distances 
from the IP.

Denote the cross sectional area of the decay volume of a 
far detector as $A$ and the length as $L$; 
the volume of the decay volume is then $V=AL$. 
If the far detector is placed at a distance $d$ from 
the IP with $d\gg L$, the probability {of the DP} to decay  
within the interval $(d,d+L)$ can be approximated by 
\be
P \simeq \exp \left[- {d \over \ell_{A'}} \right] {L \over \ell_{A'} }, 
\ee
where $\ell_{A'}$ is the decay length of the DP.
The number of DPs that 
disintegrate inside the decay volume is then given by 
\be
N \simeq N_{\rm IP} {A \over 4 \pi d^2} P 
=   N_{\rm IP} {1 \over 4 \pi } {V \over d^3} 
\exp \left[- {d \over \ell_{A'}} \right] {d \over \ell_{A'} }, 
\label{approx_nsig}
\ee
where $N_{\rm IP}$ is the total number of DPs produced at the IP, 
and we have assumed an isotropic distribution for DPs  
for simplicity. 
Thus for given $N_i$, $V$, and $d$, the optimal decay 
length to be probed is $\ell_{A'} = d$. 
Eq.\ \eqref{approx_nsig} also suggests that in order to obtain 
a large signal of LLPs, one should build a large decay volume and place 
it close to the IP if the SM backgrounds are under control; 
see also \cite{FACET:Green} for a similar discussion.

Next we compare two detectors with different $V$ and $d$. 
The ratio of the number of events {is} given by 
\be
{N_1 \over N_2} = 
{V_1 \over V_2} \left[ {d_2 \over d_1} \right]^2
\exp \left[- {d_1-d_2 \over \ell_{A'}} \right] . 
\ee
Using the parameters given in Table \ref{tab:detectors}, we find that 
${N_{\rm FACET}/N_{\rm FASER}} \simeq 7 \times 10^3 
\exp(380\, {\rm m}/  \ell_{A'})$. 
Thus the number of events in FACET is at least 
$7 \times 10^3$ times larger than FASER, 
if one neglects the background considerations and 
other effects. 
This is the main reason that the contours of FACET 
sensitivity are much larger than FASER. 
Similarly, we find that ${N_{\rm FACET}/N_{\rm FASER2}} \simeq 18 
\exp(380\, {\rm m}/  \ell_{A'})$. 
We find that these ratios between FACET and FASER(2) estimated 
here are consistent with the results from 
our simulations.\footnote{For 
example, for the model point $m_{A'} = 0.5$ GeV 
and $\epsilon_1 = 2.9 \times 10^{-7}$ in Fig.~\ref{fig:psi5DP}, 
we find that $N_{\rm FACET}/N_{\rm FASER}
\simeq 8400$ and 
$N_{\rm FACET}/N_{\rm FASER2}
\simeq 33$ in our simulations.}

\section{summary}
\label{sec:summary}

We study the capability of the various new lifetime 
frontier experiments in probing long-lived 
dark photon models. 
We consider both the minimal dark photon model, 
and the dark photon model proposed by some of us recently 
that has an enhanced long-lived dark photon signal 
at the LHC.

In the new dark photon model that has an 
{enhanced} long-lived dark photon signal at the LHC, 
the standard model is extended 
by the Stueckelberg mechanism to include 
a hidden sector, which consists of two gauge bosons
and one Dirac fermion $\psi$. 
The Stueckelberg mass terms eventually lead to 
a GeV-scale dark photon $A'$
and a TeV-scale $Z'$ 
with couplings $\epsilon_1$ and $\epsilon_2$ 
to the SM sector respectively. 
The dark photon signal at the LHC in this new model 
is enhanced because it is proportional to $\epsilon_2$ 
which can be significantly larger than $\epsilon_1$,
which is small so that the dark photon is long-lived. 
We compute various experimental constraints 
on the $\epsilon_2$ parameter including the most recent 
{constraints} on millicharge from 
the ArgoNeuT and milliQan demonstrator experiments. 
We also take into account the 
experimental constraints on the $\epsilon_1$ parameter, 
including our recasting of  
the recent LHCb model-independent limits 
on the HR process in our model.

There are three major production channels for 
the long-lived dark photon in the parameter space of interest: 
the MD, PB, and HR processes. 
The MD and PB are present in both the minimal dark photon model 
and the new dark photon model, 
and are mostly distributed in the forward region. 
The HR process, however, is only present in the new dark photon model, 
and {has} significant contributions to both the forward region 
and the transverse region 
(but still with dominant contributions in the forward region). 
We find that the HR process provides the dominant contributions 
for large dark photon mass, which opens up new 
parameter space to be probed by various new 
lifetime-frontier detectors.

We provide a mini-overview on the various  
lifetime-frontier detectors and select four detectors 
for further detailed analysis, 
which include the far detectors 
FACET, FASER (and its upgraded version, FASER2), 
and MATHUSLA, 
and the future precision timing detector CMS-MTD. 
We compute the sensitivity contours 
in the parameter 
space spanned by the dark photon mass and the 
parameter $\epsilon_1$. 
For example, with $ 300\, {\rm fb}^{-1}$ ($3\,{\rm ab}^{-1}$) data at the HL-LHC, 
FACET can probe the parameter space  
up to $ m_{A'} \simeq 1.9 \, (15)$ GeV, 
for the case where $m_\psi = 15$ GeV. 
We find that the sensitivity contours from FACET 
and  MATHUSLA are significantly enlarged by the HR process,  
and the CMS-MTD is only sensitive to the HR process. 
The enhancement in the central transverse detector MATHUSLA 
is mainly due to the fact that 
the MD and PB events are highly concentrated in the 
forward direction, and the HR process has some significant 
contributions in the transverse direction.

We further compare the signal events between the two far forward detectors: 
FACET and FASER. 
We find that FACET is likely to detect many more events than 
FASER, which is mainly due to the larger decay volume of the FACET detector 
and its smaller distance from the interaction point. 
The FASER2 detector, with a much larger decay volume 
than FASER, can somewhat 
offset the effects of the long distance from the interaction point. 
Thus we find that the FACET contours are larger than FASER and FASER2 
in our analysis.

We also find that there exists parameter space that can be probed by different kinds of 
lifetime-frontier experiments. Thus, for example, if a long-lived dark photon 
signal {were} found in one precision timing detector (e.g., CMS-MTD), 
it could then be verified by a  
far forward {detector} (e.g., FACET)  
and a far transverse {detector} (e.g., MATHUSLA).

\section{Acknowledgment}

We thank Michael Albrow for correspondence and discussions on FACET   
and for some suggestions on the FACET analysis. 
We thank Greg Landsberg for correspondence and comments 
on the millicharge constraints and on the dark photon model. 
We thank Michael William for correspondence on  
the recent LHCb model-independent analysis.
The work is supported in part  
by the National Natural Science 
Foundation of China under Grant No.\ 
11775109.

\bibliography{ref.bib}{}
\bibliographystyle{utphys28mod}

\end{document}